\documentclass[prd,10pt,nofootinbib,superscriptaddress]{revtex4-2}
\usepackage{bm}
\usepackage{graphicx}
\usepackage{hyperref}
\usepackage{comment}
\usepackage{color}
\usepackage{orcidlink}
\usepackage{ulem}
\usepackage{subfigure}
\hypersetup{colorlinks=true,linkcolor=red,citecolor=blue}
\usepackage{xcolor}
\usepackage{amsmath,amsfonts,amssymb,amstext}
\usepackage{multirow}
\usepackage{float}
\usepackage{cases}
\usepackage{epsfig}%

\begin{document}
\title{A comprehensive dynamical and phenomenological analysis of structure growth in curvature-modulated coupled quintessence scenario}
 
\author{Anirban Chatterjee \orcidlink{0000-0002-8904-7133}}
\email{Corresponding author: anirbanchatterjee@nbu.edu.cn \& iitkanirbanc@gmail.com}
\author{Yungui Gong \orcidlink{0000-0001-5065-2259}} 
\email{gongyungui@nbu.edu.cn}
\affiliation{Institute of Fundamental Physics and Quantum Technology, Department of Physics, School of Physical Science and Technology, Ningbo University, Ningbo, Zhejiang 315211, China}

\begin{abstract}  
We investigate an interacting dark energy–dark matter model within the quintessence framework, characterized by the coupling term
$Q_0 = \alpha \kappa \rho_m \dot{\phi} \left[1 - \beta R/(6H^2) \right]$, and the scalar field evolves under an exponential potential $V(\phi) = V_0 e^{-\lambda \kappa \phi}$, with parameters $\alpha$, $\lambda$, and $\beta$. Recasting the cosmological equations into a first-order autonomous system using dimensionless variables, we perform a phase space analysis to identify conditions for stable, non-phantom accelerating attractors. The Ricci scalar term, controlled by $\beta$, significantly affects the stability of critical points, with attractors transitioning to repellers for higher values of $\beta$. We also analyze linear scalar perturbations, focusing on the matter density contrast $\delta_m$ and the growth index $\gamma$. Additionally, we compute the deceleration and jerk parameters, the Hubble rate, and the distance modulus $\mu(z)$, showing good agreement with observational data. The model naturally addresses the cosmic coincidence problem through scalar field tracking behavior. For moderate parameter values, matter perturbations continue to grow into the future, capturing both background and perturbative dynamics effectively. This framework thus offers a consistent and observationally viable approach to interacting dark energy.
\end{abstract}

\maketitle

\section{Introduction}
\label{sec:Intro}

Observational evidence accumulated over the past few decades strongly indicates that the universe is undergoing a phase of accelerated expansion. This transition—from a decelerating to an accelerating regime—occurred relatively late in the cosmic timeline. The first compelling observational confirmation of this phenomenon emerged from the analysis of luminosity distance versus redshift data from Type Ia supernovae (SNe Ia) \cite{ref:Riess98, ref:Perlmutter}. Subsequent independent confirmations have come from various cosmological probes, including measurements of temperature anisotropies in the cosmic microwave background (CMB) by the WMAP mission \cite{WMAP:2003elm,Hinshaw:2008kr}, the detection of baryon acoustic oscillation (BAO) features in large-scale galaxy surveys \cite{SDSS:2005xqv}, and the study of the matter power spectrum derived from galaxy distribution data \cite{mps1,mps2}. The late-time acceleration is typically attributed to a mysterious component known as dark energy (DE), which is believed to possess negative pressure and drive the accelerated expansion. Despite its observationally inferred role, the physical nature and origin of dark energy remain among the most significant unresolved questions in modern cosmology. Alongside dark energy, which is estimated to constitute roughly 70\% of the total energy density of the universe, another substantial component is dark matter (DM). This non-luminous matter interacts primarily through gravity and is inferred from various astrophysical observations, such as the flat rotation curves of spiral galaxies \cite{Sofue:2000jx}, gravitational lensing effects \cite{Bartelmann:1999yn}, micro-lensing events, and phenomena involving colliding galaxy clusters, most notably the Bullet Cluster \cite{Clowe:2006eq}. Dark matter accounts for approximately 26\% of the cosmic energy budget, while ordinary baryonic matter contributes only about 4\%. These proportions have been precisely estimated using data from the PLANCK satellite mission \cite{Planck:2018vyg}, which provides a comprehensive picture of the universe’s large-scale structure and composition.\\

To explain the observed late-time acceleration of the universe, numerous theoretical frameworks have been proposed to model dark energy (DE). Among the most widely studied is the phenomenological \(\Lambda\)CDM model, in which Cold Dark Matter (CDM) constitutes the dominant matter component, and the accelerated expansion is attributed to the cosmological constant \(\Lambda\), introduced into Einstein's field equations via the term \(\Lambda g_{\mu\nu}\). While this model can be calibrated to match observational data, it suffers from notable theoretical challenges, including the coincidence problem \cite{Zlatev:1998tr} and the fine-tuning problem \cite{Martin:2012bt}, motivating the search for alternative DE models.  One such class comprises field-theoretic models, where the energy-momentum tensor is modified through the inclusion of an additional scalar field, independent of matter and radiation. These scalar field models aim to generate negative pressure sufficient to drive acceleration. Quintessence models \cite{Peccei:1987mm,Ford:1987de,Peebles:2002gy,Nishioka:1992sg,Ferreira:1997au,Ferreira:1997hj,Caldwell:1997ii,Carroll:1998zi,Copeland:1997et} achieve this via slowly rolling scalar field potentials, while \(k\)-essence models \cite{Fang:2014qga,ArmendarizPicon:1999rj,ArmendarizPicon:2000ah,ArmendarizPicon:2000dh,ArmendarizPicon:2005nz,Chiba:1999ka,ArkaniHamed:2003uy,Caldwell:1999ew,Bandyopadhyay:2017igc,Bandyopadhyay:2018zlz,Bandyopadhyay:2019ukl,Bandyopadhyay:2019vdd,Chatterjee:2022uyw} rely on non-canonical kinetic terms to induce acceleration. In both cases, the scalar field dynamics naturally yield the negative pressure required to explain the accelerated expansion. A different approach modifies the geometric sector of the gravitational field equations, rather than introducing new matter content. These models alter the Einstein-Hilbert action to achieve late-time acceleration and include \(f(R)\) gravity, scalar-tensor theories, Gauss-Bonnet gravity, and brane-world scenarios, as reviewed in \cite{fr1,fr2,fr3,fr4,fr5,fr6,fr7,fr8,fr9,fr10}. Such geometrically modified theories provide rich phenomenology and serve as compelling alternatives to the standard cosmological constant paradigm.\\

Numerous studies have explored the dynamics and evolution of dark energy (DE) across a variety of theoretical models, typically assuming that DE evolves independently under gravitational influence without any direct interaction with the matter sector. However, growing attention has recently been directed toward models that allow for non-gravitational interactions between dark energy and dark matter (DM). There is no fundamental constraint that forbids such interactions, provided that the total energy-momentum tensor of the universe remains conserved, \textit{i.e.,} \( \nabla_\mu T^{\mu\nu} = 0 \). This condition can either be satisfied through the independent conservation of each component or via energy-momentum exchange encoded by \( \nabla_\mu T_i^{\mu\nu} = Q_i^\nu \neq 0 \), with the constraint \( \sum_i Q_i^\nu = 0 \). The emergence of interaction terms \( Q_i^\nu \) implies a dynamical coupling between cosmic components. Interestingly, such interactions naturally arise when mapping modified gravity theories into scalar-tensor frameworks via conformal transformations \cite{mgst1,mgst2,mgst3,mgst4}. Furthermore, coupling a scalar field to the matter sector can help alleviate the cosmological coincidence problem \cite{Chimento:2003iea}. Interacting DE-DM scenarios have also been shown to influence key cosmological observables such as the cosmic microwave background (CMB) and the matter power spectrum, offering a pathway to reconcile tensions between early- and late-universe measurements. Notably, these models can address the well-known Hubble tension—namely, the \( 4 \)–\( 5\sigma \) discrepancy between the Hubble constant \( H_0 \) measured locally by the SH0ES collaboration and the value inferred from Planck CMB data \cite{ht1,ht2,An:2017crg,Pourtsidou:2016ico,Kumar:2017dnp, Yang:2018euj,DiValentino:2019ffd,DiValentino:2019jae,Nunes:2022bhn,Vagnozzi:2023nrq}. Combined analyses, such as those involving Kilo Degree Survey (KiDS) cosmic shear data and Planck angular power spectra, also suggest a preference for interacting DE-DM models over the standard \(\Lambda\)CDM framework \cite{An:2017crg,kids,plxi}.\\

This study aims to investigate how interactions between dark energy (DE) and dark matter (DM), particularly those modulated by spacetime curvature, influence the growth of cosmic structures through the gravitational evolution of scalar perturbations in a spatially flat universe. While some quantum field theory-based frameworks predict intrinsic couplings between the DE and DM sectors \cite{qft1,qft2}, our approach adopts a phenomenological model, incorporating an effective interaction term that depends on the background curvature. The analysis begins at the level of background cosmology, where the universe is described by the flat Friedmann–Lemaître–Robertson–Walker (FLRW) metric. The interaction is introduced via non-vanishing source terms in the continuity equations for both DE and DM components, ensuring total energy-momentum conservation. These source terms are typically parametrized in terms of the energy densities of the respective dark sectors. However, distinguishing among different interaction scenarios and assessing their viability requires going beyond homogeneous background dynamics. To this end, we analyze the evolution of inhomogeneities using linear cosmological perturbation theory. Assuming the absence of anisotropic stress and adopting the Newtonian gauge, scalar perturbations in the metric can be described by a single scalar potential. Since cosmic structures predominantly form on sub-horizon scales, we focus on perturbations within the matter sector, neglecting those in the DE component. The evolution of structure is captured by the DM density contrast \( \delta \), derived from the perturbed continuity equation, and the velocity divergence \( \theta \), governed by the Euler equation through momentum conservation. These perturbation variables are coupled via the Poisson equation, which emerges from the time-time component of Einstein’s field equations. In the presence of a curvature-modulated interaction—where the coupling is suppressed at early times and becomes effective in the low-curvature late universe—the impact on structure growth is expected to deviate from standard \(\Lambda\)CDM predictions, offering new signatures to probe the dark sector dynamics.\\

To describe the interaction between dark energy and dark matter in a covariant and dynamically motivated manner, we consider an interaction Lagrangian of the form \( \mathcal{L}_{\text{int}} = -\alpha \kappa \phi {\rm J} e^{-\gamma R} \), where \( \alpha \) characterizes the strength of the coupling, \( J^\mu = \rho_m u^\mu \) is the dark matter current constructed from the energy density \( \rho_m \) and the four-velocity \( u^\mu \), projected scalar current is defined as ${\rm J}=-u_{\mu}J^\mu$ and \( \gamma \) encodes the influence of curvature through the Ricci scalar \( R \). The motivation for this particular form stems from both phenomenological and theoretical considerations. First, the coupling is constructed as a scalar contraction of the scalar field \( \phi \) with the matter current \( J^\mu \), ensuring covariant consistency and minimal assumptions about the microphysics of the dark sector. Second, the inclusion of the exponential modulation \( e^{-\gamma R} \) introduces a non-linear and curvature-sensitive suppression mechanism: at early times, when spacetime curvature is large (\( R \gg 1 \)), the exponential term significantly damps the interaction, thereby preserving the success of early-universe cosmology such as Big Bang Nucleosynthesis and the formation of the Cosmic Microwave Background. At late times, as the curvature drops (\( R \to 0 \)), the interaction naturally becomes stronger, allowing for dynamical energy exchange between the dark components precisely when dark energy begins to dominate the expansion. This behavior mimics a built-in screening mechanism, where the interaction is self-regulated by the cosmic background geometry. From a dimensional perspective, since the Ricci scalar \( R \) has units of inverse length squared, the modulation parameter \( \gamma \) must carry dimensions of inverse curvature, \textit{i.e.,} \( [\gamma] = R^{-1} \), to keep the exponential term dimensionless. To facilitate cosmological analysis and embed the modulation scale in terms of the Hubble expansion, we define \( \gamma = \beta / 6H^2 \), where \( H \) is the Hubble parameter and \( \beta \) is a dimensionless parameter that serves as the primary model variable in our study. This substitution allows the curvature suppression to evolve dynamically with the expansion of the universe, while maintaining dimensional consistency. Moreover, the exponential form avoids inverse powers of curvature (which can be problematic or lead to instabilities) and instead offers a smooth and well-controlled transition from negligible to significant coupling. This structure is also well-motivated within effective field theory (EFT) approaches to dark sector interactions. Thus, the chosen interaction Lagrangian provides a physically grounded and technically robust framework for exploring energy exchange between dark energy and dark matter, while ensuring compatibility with both cosmological observations and theoretical consistency.\\

Unlike most existing studies that introduce the interaction term phenomenologically at the level of the continuity equations, thereby modifying them into non-conserving forms, our approach derives the interaction directly from the Lagrangian. This ensures a covariant and variationally consistent formulation of the dynamics. While earlier works \cite{Wetterich:1994bg, Amendola:1999er, Bernardi:2016xmb, Duniya:2013eta}have considered forms such as \( Q_0 \sim \rho_m \dot{\phi} \) \cite{Wetterich:1994bg, Amendola:1999er}, or extended expressions like \( Q_0 \sim (\rho_m + \rho_\phi)\dot{\phi} \) \cite{Bernardi:2016xmb}, these often lack theoretical grounding or fail to remain viable across all cosmological epochs. Moreover, purely dark energy-based couplings such as \( Q_0 \sim \rho_\phi \dot{\phi} \) \cite{Duniya:2013eta} are generally ineffective in influencing structure formation due to the non-clustering nature of dark energy at sub-horizon scales. The choice of that particular interacting Lagrangian leads to an interaction term that is well-motivated from a field-theoretical perspective, taking the form $Q_0 = \alpha \kappa \rho_m \dot{\phi} \left(1 - \frac{\beta R}{6H^2} \right),$ where the curvature-modulated factor \( \left(\frac{\beta R}{6H^2} \right) \) dynamically regulates the energy exchange between dark matter and dark energy. By formulating the interaction from the Lagrangian level, we not only capture these physical features consistently but also ensure that the resulting energy exchange terms are grounded in field-theoretic principles rather than introduced by hand. This offers a more robust and unified framework for analyzing interacting dark energy--dark matter models, particularly in the context of structure formation, background evolution, and cosmic acceleration. Within the framework of quintessence models, late-time acceleration is typically realized when the scalar field evolves slowly, such that the kinetic energy \( \dot{\phi}^2 \) is subdominant to the potential energy \( V(\phi) \), ensuring \( \omega_\phi < -1/3 \). However, during earlier epochs—particularly during matter-dominated phases relevant for structure formation—the scalar field kinetic term \( \dot{\phi} \) cannot be neglected. Expression for \( Q_0 \) thus captures a scenario in which the interaction rate is modulated both by the dark matter energy density and by two key temporal scales: the field evolution rate \( \dot{\phi} \) and the Hubble expansion rate \( H \). Importantly, the curvature-modulating factor \( \left(  \frac{\beta R}{6H^2}\right) \) introduces a dynamic suppression mechanism: the interaction is negligible at early times when curvature \( R \sim H^2 \) is large, but becomes significant at late times as \( R \) decreases, thereby enabling a self-regulating energy transfer. We study the cosmological implications of this curvature-modulated interaction in the context of linear perturbation theory, focusing on the growth of matter fluctuations. To systematically assess the background and perturbative behavior, we employ the dynamical system analysis (DSA) technique, which allows us to characterize fixed points, assess their stability, and compare the structure growth behavior across different interacting models \cite{Marcondes:2016reb,Dutta:2017wfd,Chatterjee:2024duy,Khyllep:2021wjd}. This approach enables a unified framework to probe both the background dynamics and the impact of interactions on the evolution of perturbations, while highlighting the unique signatures introduced by curvature-modulated coupling.  \\

Incorporating interactions between dark energy (DE) and dark matter (DM) adds substantial complexity to the evolution equations governing both the background and perturbation levels of the universe. These complexities make it increasingly challenging to extract analytic or even stable numerical solutions. In such scenarios, dynamical system analysis (DSA) emerges as a powerful mathematical framework for obtaining global qualitative insights into the system's behavior without requiring explicit solutions \cite{Marcondes:2016reb,Dutta:2017wfd,Chatterjee:2024duy,Khyllep:2021wjd,Chatterjee:2021ijw,Chatterjee:2021hhj,Hussain:2022osn,Bhattacharya:2022wzu,Hussain:2023kwk,Chatterjee:2023uga,Cabral:2009hoy,Tsujikawa:2012hv}. This approach reformulates the cosmological evolution equations into an autonomous system of first-order differential equations in terms of suitably defined dimensionless variables. In the case of our model, where the interaction term is given by \( Q_0 = \alpha \kappa \rho_m \dot{\phi} \left(1 - \frac{\beta R}{6H^2} \right) \), the dynamical system is three-dimensional and constructed in terms of the variables \( x, y, u \), evolved with respect to the e-folding number \( N = \ln a \). Where, $x$ and $ y$ represent the variables constructed from the kinetic and potential terms of the scalar field and  \( u \) effectively tracks the growth rate of matter perturbations. A detailed phase space analysis of the system reveals the nature and stability of critical points, which depend on the model parameters: the DE-DM coupling strength \( \alpha \) and the potential slope \( \lambda \) in the exponential quintessence potential \( V(\phi) = V_0 e^{-\lambda \kappa \phi} \). Each critical point corresponds to a distinct cosmological phase, whose characteristics—such as matter and DE density parameters, the effective equation of state, and the nature of perturbation growth—help in understanding the underlying dynamics, including how the model can address the cosmic coincidence problem. Notably, the presence of interaction modifies the growth history of structures and can lead to prolonged or delayed growth of perturbations in the late universe compared to non-interacting models. This is particularly relevant when studying the evolution of the density contrast \( \delta_m \), the growth rate \( u \), and the growth index \( \gamma \), all of which are sensitive to the coupling parameter \( \alpha \). From an observational standpoint \cite{Scolnic:2021amr, DES:2024jxu, Gao:2024ily,Gadbail:2024rpp, DESI:2024mwx, Chatterjee:2024rbh}, we test the viability of the interacting model by comparing its predictions against key cosmological observables. This includes analyzing the distance modulus \( \mu(z) \) versus redshift \( z \) using the Pantheon+ SNe Ia dataset, and the Hubble parameter evolution \( H(z) \) compared to observational \( H(z) \) data points and the standard \(\Lambda\)CDM model. Furthermore, we study the cosmographic evolution—particularly quantities like the deceleration and jerk parameters—as another diagnostic to contrast the interacting model’s predictions with those of \(\Lambda\)CDM. Overall, this unified DSA framework not only facilitates a clear identification of viable attractor solutions and growing-mode trajectories independent of initial conditions, but also enables a comprehensive understanding of the coupled dynamics at both background and perturbation levels, with direct relevance to current and future observational constraints.\\

The structure of this article is as follows. In Sec.~\ref{sec:IDS}, we formulate the field equations governing a general interacting field-fluid system, covering both the background dynamics and the evolution of linear matter perturbations. Section~\ref{sec:DSA} is dedicated to constructing the corresponding dynamical system, where we define and recast key cosmological quantities in terms of suitable dimensionless dynamical variables. In Sec.~\ref{sec:ROD}, we present a summary of the results from the dynamical stability analysis, highlighting the phase space trajectories and the qualitative behavior of the interacting system. Lastly, Sec.~\ref{con} provides our concluding remarks, summarizing the key insights drawn from this study.

  \section{Theoretical Framework: Evolution at the Level of Background and Perturbations in an Interacting DE-DM Scenario}
\label{sec:IDS}

\subsection{Grand action and  Motivation}

We consider an interacting dark energy–dark matter model in a spatially flat FLRW background, where the interaction arises through an additive term in the total Lagrangian. The total action is given by
\begin{equation}
S = \int d^4x \sqrt{-g} \left( \mathcal{L}_{\text{grav}} + \mathcal{L}_\phi + \mathcal{L}_m + \mathcal{L}_{\text{int}} \right)
\end{equation}
where each term corresponds to a distinct physical sector. The gravitational part is governed by the Einstein-Hilbert term \( \mathcal{L}_{\text{grav}} = R/(2\kappa^2) \), with \( \kappa^2 = 8\pi G \) and \( R \) the Ricci scalar. The dark energy sector is described by a canonical scalar field \( \phi \) with Lagrangian $\mathcal{L}_\phi = \frac{1}{2} g^{\mu\nu} \partial_\mu \phi \partial_\nu \phi + V(\phi)$, where the potential takes the exponential form \( V(\phi) = V_0 e^{-\lambda \kappa \phi} \). The energy-momentum tensor associated with the scalar field is derived by variation with respect to the metric and is given by $T^{(\phi)}_{\mu\nu} = \partial_\mu \phi \partial_\nu \phi - g_{\mu\nu} \left( \frac{1}{2} g^{\alpha\beta} \partial_\alpha \phi \partial_\beta \phi + V(\phi) \right)$.  In the FLRW background, this yields the scalar field energy density and pressure as, $\rho_\phi = \frac{1}{2} \dot{\phi}^2 + V(\phi), \quad p_\phi = \frac{1}{2} \dot{\phi}^2 - V(\phi)$. The dark matter sector is modeled as a pressureless perfect fluid described by \( \mathcal{L}_m \), with energy density \( \rho_m \) and four-velocity \( u^\mu \). We propose a covariant interaction Lagrangian of the form \( \mathcal{L}_{\text{int}} = -\alpha \kappa \phi {\rm J} e^{-\gamma R} \) to model energy exchange between dark energy and dark matter, where \( \gamma \) modulates curvature effects via the Ricci scalar \( R \). The exponential term ensures that the interaction is naturally suppressed at early times (high curvature) and becomes significant at late times (low curvature), acting as a dynamical screening mechanism. Dimensional consistency requires \( [\gamma] = R^{-1} \), and later we will define \( \gamma = \beta / 6H^2 \), introducing a dimensionless parameter \( \beta \) that governs the strength of curvature sensitivity. This formulation offers a smooth and stable evolution of the coupling and is well-motivated by effective field theory considerations, providing a robust framework for addressing late-time cosmic acceleration and the dark sector coupling in a geometrically consistent way.

 \subsection{Energy-Momentum Conservation and Interaction Source Term}

The total energy-momentum tensor is given by the sum of the scalar field and matter contributions:
\begin{equation}
T_{\mu\nu} = T^{(\phi)}_{\mu\nu} + T^{(m)}_{\mu\nu}
\end{equation}
While the Bianchi identity \( \nabla^\mu G_{\mu\nu} = 0 \) ensures total energy-momentum conservation \( \nabla^\mu T_{\mu\nu} = 0 \), the presence of an interaction between dark matter and dark energy leads to individual non-conservation of each component:
\begin{equation}
\nabla^\mu T^{(m)}_{\mu\nu} = Q_\nu, \quad \nabla^\mu T^{(\phi)}_{\mu\nu} = -Q_\nu
\end{equation}
where \( Q_\nu \) characterizes the four-vector of energy and momentum transfer between the dark sectors. As we consider an interaction term of the form $S_{\text{int}} = -\int d^4x \sqrt{-g} \, \alpha \kappa \phi {\rm J} e^{-\gamma R}$ with \({\rm J} = -u_{\mu} J^\mu = \rho_m \) representing the projected scalar current. Varying this action with respect to the metric and scalar field leads to an interaction source term and working in the no-momentum-transfer frame gives the form $Q_\nu = -\alpha \kappa \rho_m u_\nu \dot{\phi} e^{-\gamma R}$. More generally, the interaction vector includes both \( \nabla_\nu \phi \) and \( \nabla_\nu R \) terms due to curvature sensitivity in the exponential. However, in a homogeneous and isotropic FLRW background, the Ricci scalar \( R \) depends only on time, and its spatial derivatives vanish. Moreover, the term proportional to \( \phi \nabla_\nu R \) is suppressed when the curvature varies slowly (\(|\dot R|/(HR)\ll 1\)) and \( \gamma \) varies mildly. Since \( \nabla_\nu R \) contributes only subdominantly in the smooth, late-time universe, we neglect higher-order curvature corrections at the background level, retaining only the dominant term aligned with the matter four-velocity. In the FLRW background with \( u^\mu = (1, 0, 0, 0) \), the interaction reduces to \( Q_0 = \alpha \kappa \rho_m \dot{\phi} \, e^{-\gamma R} \). Approximating \( e^{-\gamma R} \simeq 1 - \gamma R \) under \( |\gamma R| \ll 1 \), valid at late times, yields a regular leading-order form. For dimensional clarity one may parameterize \( \gamma = \beta/k^{2} \) with a fixed mass scale \(k\); evaluated on the background one can take \(k^{2} \simeq 6H^{2}\), so that \( \gamma R \simeq \beta R/(6H^{2}) \), with \( \beta \) controlling the curvature sensitivity. This leads to a dynamical screening mechanism where the interaction is suppressed in the early universe and activated at late times. The resulting expression takes the form
\begin{equation}
Q_0 = \alpha \kappa \rho_m \dot{\phi} \left(1 - \frac{\beta R}{6H^2} \right),
\end{equation}
which is regular, avoids higher-derivative instabilities, and remains theoretically consistent within the effective field theory (EFT) approach.

\subsection{Modified Continuity Equations and Friedmann Dynamics}

The total conservation equation \( \nabla^\mu T_{\mu\nu} = 0 \) leads to the split:
\begin{align}
\dot{\rho}_m + 3H \rho_m &= Q_0, \\
\dot{\rho}_\phi + 3H(\rho_\phi + p_\phi) &= -Q_0
\label{eq:cont}
\end{align}
indicating energy transfer from one sector to the other depending on the sign of \( Q_0 \).\\

The Einstein field equations give the standard Friedmann equations:
\begin{align}
3H^2 &= \kappa^2 (\rho_m + \rho_\phi), \\
2\dot{H} + 3H^2 &= -\kappa^2 p_\phi.
\end{align}

The dimensionless density parameters are defined as
\begin{equation}
\Omega_m = \frac{\kappa^2 \rho_m}{3H^2}, \quad \Omega_\phi = \frac{\kappa^2 \rho_\phi}{3H^2},
\end{equation}
satisfying the closure relation
\begin{equation}
\Omega_m + \Omega_\phi = 1.
\end{equation}

Finally, the deceleration parameter and total equation of state are given by
\begin{equation}
q = -\frac{\ddot{a}}{aH^2}, \quad \omega_{\text{tot}} = \frac{p_\phi}{\rho_m + \rho_\phi} = \frac{2q - 1}{3}.
\end{equation}

The accelerated expansion of the universe occurs when the deceleration parameter satisfies \( q < 0 \), or equivalently, when the total equation of state parameter satisfies \( \omega_{\text{tot}} < -\frac{1}{3} \).

\subsection{Linear Perturbations in Newtonian Gauge}

In the Newtonian gauge, scalar perturbations are considered in the line element:
\begin{equation}
ds^2 = a^2(\tau) \left[ -(1 + 2\Psi) d\tau^2 + (1 - 2\Phi) \delta_{ij} dx^i dx^j \right],
\end{equation}
where $\tau$ is the conformal time. Assuming no anisotropic stress, we take $\Psi = \Phi$.

Perturbing the energy-momentum tensor of the matter component yields
\begin{equation}
\delta T^0_{;0} = -\delta \rho_m, \quad \delta T^i_{;0} = -\frac{1}{a} \rho_m \delta u^i.
\end{equation}
The perturbed interaction current is written as $Q_\nu = Q_\nu + \delta Q_\nu$.

The perturbation equations in Fourier space are given by:
\begin{align}
-\delta'_m + \frac{Q_0}{\rho_m} \delta_m - \theta + 3\Phi' = \frac{\delta Q_0}{\rho_m}, \
\theta' + \left( \mathcal{H} - \frac{Q_0}{\rho_m} \right) \theta - k^2 \Phi = \frac{i k^i \delta Q_i}{\rho_m},
\end{align}
where $\theta = a^{-1} ik^j \delta u_j$ is the velocity divergence and $\mathcal{H} = a'/a$.

The Poisson equation reads:
\begin{equation}
k^2 \Phi = -\frac{3}{2} \mathcal{H}^2 \Omega_m \delta_m.
\end{equation}

Combining the above, the second-order evolution equation for the matter density contrast becomes:
\begin{equation}
\delta''_m - (\mathcal{Q} - \mathcal{K}) \delta'_m - \left( \frac{3}{2} \mathcal{H}^2 \Omega_m + \mathcal{Q}' + \mathcal{K} \mathcal{Q} \right) \delta_m = - \frac{i k^i \delta Q_i}{\rho_m},
\label{eq:fullSOE}
\end{equation}
with
\begin{equation}
\mathcal{Q} = \frac{Q_0}{\rho_m} - \frac{\delta Q}{\rho_m \delta_m}, \quad \mathcal{K} = \mathcal{H} - \frac{Q_0}{\rho_m}.
\end{equation}

In the absence of any interaction ($Q_0 = 0$, $\delta Q = 0$), the standard equation for linear growth is recovered:
\begin{equation}
\delta''_m + \mathcal{H} \delta'_m - \frac{3}{2} \mathcal{H}^2 \Omega_m \delta_m = 0.
\end{equation}
 
Expressed in terms of cosmic time \( t \) rather than conformal time \( \tau \), the perturbation equation becomes
\begin{equation}
\ddot{\delta}_m + 2 H \dot{\delta}_m - \frac{3}{2} H^2 \Omega_m \delta_m = 0.
\end{equation}

\section{Dynamical Stability Analysis of the Interacting System}
\label{sec:DSA}

The evolution equations governing the universe, especially in the presence of dark energy-dark matter (DE-DM) interactions, are generally complex, making analytical treatment difficult. To study the qualitative behavior of cosmological evolution, we adopt a dynamical systems approach by reformulating the system as a set of first-order autonomous differential equations using suitable dimensionless variables. This framework enables a deeper understanding of both the background dynamics and the growth of linear perturbations, without relying solely on numerical integration. We consider a quintessence-type scalar field potential of the form $V(\phi) = V_0 e^{-\lambda \kappa \phi}$, where $\lambda$ is a dimensionless slope parameter. The interaction between the dark sectors is characterized by the source term, $Q_0 = \alpha \kappa \rho_m \dot{\phi} \left(1 - \frac{\beta R}{6H^2}  \right)$, where $\alpha$ and $\beta$ are coupling constants. The sign of $\alpha$ determines the energy transfer direction for both the sectors. This choice of interaction is motivated by effective field theory considerations that ensure covariant coupling while avoiding higher-derivative instabilities. The Ricci scalar modulation introduces a natural time dependence, suppressing the interaction during early epochs (high curvature) and enhancing it at late times (low curvature), which effectively mimics a screening mechanism and is consistent with observational constraints.\\

\subsection{Construction of 3-D dynamical system}

Substituting the expression for energy density ($\rho_\phi$) into the scalar field continuity eq. (\ref{eq:cont}), the evolution equation for $\phi$ under the curvature-modulated interaction becomes:
\begin{equation}
\ddot{\phi} + 3H \dot{\phi} + \frac{dV}{d\phi} = - \frac{Q_0}{\dot{\phi}} 
\label{eq:D2new}
\end{equation}
This above equation illustrates how the background evolution of the scalar field is modified by the presence of the dark matter current and the Ricci scalar, leading to non-trivial feedback effects on both expansion history and the development of large-scale structures. The coupling term thus not only enables energy exchange between the sectors but also enhances the dynamical richness of perturbation evolution, offering a framework capable of addressing the cosmic coincidence problem.\\

To study the dynamics, we introduce the following dimensionless variables:
\begin{equation}
x = \frac{\kappa \dot{\phi}}{\sqrt{6} H}, \qquad
y = \frac{\kappa \sqrt{V}}{\sqrt{6} H}, \qquad
u = \frac{d \ln \delta_m}{d \ln a},
\end{equation}
where $x$ and $y$ characterize the scalar field kinetic and potential energy densities relative to the Hubble expansion, and $u$ quantifies the growth rate of the matter density contrast $\delta_m$. The potential parameter is given by $\lambda = -\tfrac{(dV/d\phi)}{\kappa V}$.\\
 
Autonomous equations of this system take the following form; 
\begin{eqnarray}
\frac{dx}{dN} &=& -3x + \sqrt{\frac{3}{2}}\,\lambda\, y^2 + \frac{3}{2}x(1 + x^2 - y^2) 
- \frac{3\alpha}{\sqrt{6}} (1 - x^2 - y^2)\left(1 - \frac{\beta}{2}(1 - 3x^2 + 3y^2)\right) \label{eq:dx}\\
\frac{dy}{dN} &=& -\frac{\sqrt{6}}{2} \lambda\, x y + \frac{3}{2} y (1 + x^2 - y^2)\label{eq:dy}\\
\frac{du}{dN} &=& \frac{3}{2}(1 - x^2 - y^2) - u^2 - \left( \frac{1}{2} - \frac{3}{2}(x^2 - y^2) 
- \alpha\left( \sqrt{6}x - \frac{\beta}{2}(1 - 3x^2 + 3y^2) \right) \right) u \label{eq:du}
\end{eqnarray}

The evolution equation for matter density perturbations Eq.~(\ref{eq:fullSOE}) in terms of $x$, $y$, and $ u$ is:
\begin{equation}
\delta''_m + \delta'_m \left[2 - \alpha \left(\sqrt{6} x - \frac{\beta}{2}(1 - 3x^2 + 3y^2)\right) \right] - \frac{3}{2} (1 - x^2 - y^2) \delta_m = 0
\label{eq:dlm}
\end{equation}
where $u = \delta'_m / \delta_m$, and positive (negative) $u$ indicates growth (decay) of density perturbations.\\

The background cosmological parameters in terms of the dynamical variables are given by;
\begin{align}
\Omega_\phi &= x^2 + y^2,   \quad & \omega_{\text{tot}} &= x^2 - y^2, \label{eq:Omega_phi_omega_tot} \\
\Omega_m &= 1 - x^2 - y^2, & q &= \frac{1}{2} + \frac{3}{2}(x^2 - y^2), \label{eq:Omega_m_q} \\
r_{\rm mc} &= \frac{\Omega_m}{\Omega_\phi}, & j &= q(1 + 2q) + 3 (x x' - y y') \label{eq:r_mc_j}.
\end{align}

The jerk parameter $j$, which represents the third derivative of the scale factor, provides insights into the rate of change of the cosmic acceleration. It is sensitive to higher-order deviations from $\Lambda$CDM and is particularly useful for distinguishing between different dynamical models of dark energy. The evolution of $j$ is influenced by both the scalar field dynamics and the curvature-modulated interaction.  Together, these expressions characterize the full dynamical structure of the coupled DE--DM system and enable a unified treatment of both expansion history and structure formation. In particular, the variables $x$ and $y$ govern the background evolution, while $u$ tracks the growth of inhomogeneities. Their combined behavior provides powerful diagnostics for the viability of interacting dark energy scenarios and the resolution of the cosmic coincidence problem.

\subsection{Critical points and their properties}

In this section, we present and analyze the results of the dynamical system associated with the curvature-modulated dark energy--dark matter interaction model introduced earlier. The system of autonomous equations derived in the previous section admits a range of critical points, each corresponding to distinct cosmological behaviors. These points are determined by setting the right-hand sides of the dynamical equations to zero and solving for $(x, y, u)$. The stability of each critical point is evaluated through the eigenvalues of the Jacobian matrix of the system, providing insight into the late-time attractors and the evolution of matter perturbations. A physically relevant and stable critical point with $u > 0$ indicates sustained growth of matter perturbations, while $u < 0$ implies decay of perturbations and asymptotic stability against structure formation. A point with $u = 0$ denotes the freezing of perturbations to a constant value. This allows us to classify the viability of various cosmological scenarios encoded in the model.

\begin{table}[H]
\centering
\renewcommand{\arraystretch}{1.5}
\begin{tabular}{|c|c|c|c|}
\hline
\textbf{CP} & $\textbf{x}$ & $\textbf{y}$ & $\textbf{u}$ \\
\hline
$A_{\pm}$ & $\pm 1$ & $0$ & $0$ \\
$B_{\pm}$ & $ \pm 1$ & $0$ & $1 + \alpha (\pm \sqrt{6} + \beta)$ \\
$C_{\pm}$ & $-\dfrac{1 + \sqrt{1 + 2 \alpha^2 (-2 + \beta) \beta}}{\sqrt{6} \alpha \beta}$ & $0$ & See explicit form below \\
$D_{\pm}$ & $\dfrac{-1 + \sqrt{1 + 2 \alpha^2 (-2 + \beta) \beta}}{\sqrt{6} \alpha \beta}$ & $0$ & See explicit form below \\
$E_{\pm}$ & $\dfrac{\lambda}{\sqrt{6}}$ & $ \pm \sqrt{1 - \lambda^2/6}$ & $0$ \\
$F_{\pm}$ & $\dfrac{\lambda}{\sqrt{6}}$ & $ {\pm} \sqrt{1 - \lambda^2/6}$ & $\frac{1}{2}(-4 + 2\alpha\lambda + \lambda^2 + \alpha\beta(-4 + \lambda^2))$ \\
$G_{\pm}$ & See explicit form below & See explicit form below & See explicit form below \\
$H_{\pm}$ & See explicit form below & See explicit form below & See explicit form below \\
$I_{\pm}$ & See explicit form below & See explicit form below & See explicit form below \\
$J_{\pm}$ & See explicit form below & See explicit form below & See explicit form below \\
\hline
\end{tabular}
\caption{Complete list of all 10 sets of critical points along with their coordinates \( (x, y, u) \).  Explicit expressions for nontrivial forms are provided below.}
\label{tab:1}
\end{table} 

In this framework, we identify up to 20 critical points depending on the specific choices of the model parameters $\alpha$, $\lambda$, and $\beta$. These include points labeled as $A_\pm$, $B_\pm$, $C_\pm$, $D_\pm$, $E_\pm$, $F_\pm$, $G_\pm$, $H_\pm$, $I_\pm$, and $J_\pm$. Not all of these points are guaranteed to exist simultaneously: the presence and reality of any given critical point are subject to specific constraints on the parameter space, such as existence conditions, boundedness of the coordinates, and physical relevance. Interestingly, many of these critical points occur in pairs that share the same background coordinates $(x, y)$ but differ in their perturbation behavior via $u$. This reflects a generic property of our chosen dynamical variable set, where the $(x, y)$ coordinates capture background expansion while $u$ encapsulates perturbation growth.  A complete classification of the fixed points, including their stability properties, parameter constraints, and associated physical quantities such as the effective equation of state and matter density, is summarized in tabs.~\ref{tab:1} and \ref{tab:2}. These results form the basis for a deeper understanding of how interactions in the dark sector can influence both the background expansion and structure formation history of the Universe.\\

 \noindent
\textbf{\underline{Explicit Coordinates for Critical Points:}}
{\scriptsize
\begin{eqnarray}
x_{C_\pm} &=& -\frac{1 + \sqrt{1 + 2 \alpha^2 (-2 + \beta) \beta}}{\sqrt{6} \alpha \beta}, \quad y_{C_\pm} = 0, \nonumber
\end{eqnarray}

\begin{align*}
u_{C_\pm} = \frac{1}{4 \alpha^2 \beta^2} \bigg( 1 + \alpha \beta - 4 \alpha^2 \beta - 2 \alpha^3 \beta^2  
 + \sqrt{1 + 2 \alpha^2 (-2 + \beta) \beta} 
 + \alpha \beta \sqrt{1 + 2 \alpha^2 (-2 + \beta) \beta} \\
 - 2 \alpha^2 \beta \sqrt{1 + 2 \alpha^2 (-2 + \beta) \beta}  \mp \alpha^2 \beta^2 \sqrt{ \frac{ \mathcal{N}_C }{ \alpha^4 \beta^4 } } \bigg),
\end{align*}

\begin{align*}
\mathcal{N}_C &= -8 \alpha^2 \beta^2 \left( 1 - 2 \alpha^2 \beta (1 + \beta) + \sqrt{1 + 2 \alpha^2 (-2 + \beta) \beta} \right) \\
&\quad + \left( 1 + \sqrt{1 + 2 \alpha^2 (-2 + \beta) \beta} + \alpha \beta \left( 1 + \sqrt{1 + 2 \alpha^2 (-2 + \beta) \beta} \right. \right. \\
&\quad \left. \left. - 2 \alpha (2 + \alpha \beta + \sqrt{1 + 2 \alpha^2 (-2 + \beta) \beta}) \right) \right)^2\\
\end{align*}

\begin{align*}
x_{D_\pm} = \frac{-1 + \sqrt{1 + 2 \alpha^2 (-2 + \beta) \beta}}{\sqrt{6} \alpha \beta} \quad
y_{D_\pm} = 0, 
\end{align*}

\begin{align*}
u_{D_\pm} &= \frac{1}{4 \alpha^2 \beta^2} \bigg( 1 + \alpha \beta - 4 \alpha^2 \beta - 2 \alpha^3 \beta^2  \mp \sqrt{1 + 2 \alpha^2 (-2 + \beta) \beta}   \mp \alpha \beta \sqrt{1 + 2 \alpha^2 (-2 + \beta) \beta} \\
&\quad \pm 2 \alpha^2 \beta \sqrt{1 + 2 \alpha^2 (-2 + \beta) \beta}   + \alpha^2 \beta^2 \sqrt{ \frac{ \mathcal{N}_D }{ \alpha^4 \beta^4 } } \bigg),
\end{align*}

\begin{align*}
\mathcal{N}_D &= 8 \alpha^2 \beta^2 \left( -1 + 2 \alpha^2 \beta (1 + \beta) + \sqrt{1 + 2 \alpha^2 (-2 + \beta) \beta} \right) \\
&\quad + \left( -1 + \sqrt{1 + 2 \alpha^2 (-2 + \beta) \beta} + \alpha \beta \left( -1 + \sqrt{1 + 2 \alpha^2 (-2 + \beta) \beta} \right. \right. \\
&\quad \left. \left. + 2 \alpha (2 + \alpha \beta - \sqrt{1 + 2 \alpha^2 (-2 + \beta) \beta}) \right) \right)^2
\end{align*}

\begin{align*}
x_{G_\pm} &=
\frac{
\alpha (-1 + 2 \beta) - \lambda + 
\sqrt{\alpha^2 (1 - 2 \beta)^2 + 
2 \alpha (1 + \beta) \lambda + 
\lambda^2}
}{
\sqrt{6} \alpha \beta \lambda
}
\\[1.5em]
y_{G_\pm} &=
-\frac{1}{\sqrt{3}} \sqrt{\Big{(}
\frac{
\alpha^2 ((1 - 2 \beta)^2 + \beta (1 + \beta) \lambda^2)
+ \lambda (\lambda - 
\sqrt{
\alpha^2 (1 - 2 \beta)^2 + 2 \alpha (1 + \beta) \lambda + \lambda^2
})
}{
\alpha^2 \beta^2 \lambda^2
}
} \\
&\quad + \frac{\alpha}{\sqrt{3} \alpha^2 \beta^2 \lambda^2}
\left[
(-2 + \beta) \lambda - \beta \lambda^3 +
(1 - 2 \beta) \sqrt{
\alpha^2 (1 - 2 \beta)^2 + 2 \alpha (1 + \beta) \lambda + \lambda^2
} \right. \\
&\quad\quad \left. +
\beta \lambda^2 \sqrt{
\alpha^2 (1 - 2 \beta)^2 + 2 \alpha (1 + \beta) \lambda + \lambda^2
}\Big{)}
\right]
\\[1.5em]
u_{G_\pm} &=
\pm \frac{1}{4 \alpha \beta \lambda}
\Bigg[
\lambda^2 - 2 \alpha \sqrt{
\alpha^2 (1 - 2 \beta)^2 + 2 \alpha (1 + \beta) \lambda + \lambda^2
}
\\
&\quad - \lambda \sqrt{
\alpha^2 (1 - 2 \beta)^2 + 2 \alpha (1 + \beta) \lambda + \lambda^2
}
- \alpha \beta \lambda \sqrt{
\alpha^2 (1 - 2 \beta)^2 + 2 \alpha (1 + \beta) \lambda + \lambda^2
}
\\
&\quad + \alpha \lambda (3 + \beta (2 + \lambda)) + 
\alpha^2 \left(2 + \beta(-4 + \lambda + 2 \beta \lambda)\right)
\\
&\quad + \alpha \beta \lambda \sqrt{
\frac{1}{\alpha^2 \beta^2 \lambda^2}
\left(
8 (\alpha + \lambda + \alpha \beta (-2 + \lambda^2)
- \sqrt{
\alpha^2 (1 - 2 \beta)^2 + 2 \alpha (1 + \beta) \lambda + \lambda^2
})
\right.
}
\\
&\quad\quad \times (\alpha (-1 + 2 \beta) - \lambda + 
\sqrt{
\alpha^2 (1 - 2 \beta)^2 + 2 \alpha (1 + \beta) \lambda + \lambda^2
})
\\
&\quad\quad + \alpha^2 (2 + \beta(-4 + \lambda + 2 \beta \lambda)) +
\lambda (\lambda - \sqrt{
\alpha^2 (1 - 2 \beta)^2 + 2 \alpha (1 + \beta) \lambda + \lambda^2
})
\\
&\quad\quad \left. +
\alpha \left(
-2 \sqrt{
\alpha^2 (1 - 2 \beta)^2 + 2 \alpha (1 + \beta) \lambda + \lambda^2
}
+ \lambda (3 + \beta (2 + \lambda -
\sqrt{
\alpha^2 (1 - 2 \beta)^2 + 2 \alpha (1 + \beta) \lambda + \lambda^2
}))
\right)
\right)^2
\Bigg]
\end{align*}
 
\begin{align*}
x_{H_\pm} &=
\frac{
\alpha (-1 + 2 \beta) - \lambda + 
\sqrt{\alpha^2 (1 - 2 \beta)^2 + 
2 \alpha (1 + \beta) \lambda + 
\lambda^2}
}{
\sqrt{6} \alpha \beta \lambda
}
\\[1.5em]
y_{H_\pm} &=
\frac{1}{\sqrt{3}} \sqrt{\Big{(}
\frac{
\alpha^2 ((1 - 2 \beta)^2 + \beta (1 + \beta) \lambda^2)
+ \lambda (\lambda - 
\sqrt{
\alpha^2 (1 - 2 \beta)^2 + 2 \alpha (1 + \beta) \lambda + \lambda^2
})
}{
\alpha^2 \beta^2 \lambda^2
}
} \\
&\quad - \frac{\alpha}{\sqrt{3} \alpha^2 \beta^2 \lambda^2}
\left[
(-2 + \beta) \lambda - \beta \lambda^3 +
(1 - 2 \beta) \sqrt{
\alpha^2 (1 - 2 \beta)^2 + 2 \alpha (1 + \beta) \lambda + \lambda^2
} \right. \\
&\quad\quad \left. +
\beta \lambda^2 \sqrt{
\alpha^2 (1 - 2 \beta)^2 + 2 \alpha (1 + \beta) \lambda + \lambda^2
}\Big{)}
\right]
\\[1.5em]
u_{H_\pm} &=
\pm \frac{1}{4 \alpha \beta \lambda}
\Bigg[
\pm \lambda^2 \mp 2 \alpha \sqrt{
\alpha^2 (1 - 2 \beta)^2 + 2 \alpha (1 + \beta) \lambda + \lambda^2
}
\\
&\quad \mp \lambda \sqrt{
\alpha^2 (1 - 2 \beta)^2 + 2 \alpha (1 + \beta) \lambda + \lambda^2
}
- \alpha \beta \lambda \sqrt{
\alpha^2 (1 - 2 \beta)^2 + 2 \alpha (1 + \beta) \lambda + \lambda^2
}
\\
&\quad + \alpha \lambda (3 + \beta (2 + \lambda)) + 
\alpha^2 \left(2 + \beta(-4 + \lambda + 2 \beta \lambda)\right)
\\
&\quad + \alpha \beta \lambda \sqrt{
\frac{1}{\alpha^2 \beta^2 \lambda^2}
\left(
8 (\alpha + \lambda + \alpha \beta (-2 + \lambda^2)
- \sqrt{
\alpha^2 (1 - 2 \beta)^2 + 2 \alpha (1 + \beta) \lambda + \lambda^2
})
\right.
}
\\
&\quad\quad \times (\alpha (-1 + 2 \beta) - \lambda + 
\sqrt{
\alpha^2 (1 - 2 \beta)^2 + 2 \alpha (1 + \beta) \lambda + \lambda^2
})
\\
&\quad\quad + \alpha^2 (2 + \beta(-4 + \lambda + 2 \beta \lambda)) +
\lambda (\lambda - \sqrt{
\alpha^2 (1 - 2 \beta)^2 + 2 \alpha (1 + \beta) \lambda + \lambda^2
})
\\
&\quad\quad \left. +
\alpha \left(
-2 \sqrt{
\alpha^2 (1 - 2 \beta)^2 + 2 \alpha (1 + \beta) \lambda + \lambda^2
}
+ \lambda (3 + \beta (2 + \lambda -
\sqrt{
\alpha^2 (1 - 2 \beta)^2 + 2 \alpha (1 + \beta) \lambda + \lambda^2
}))
\right)
\right)^2
\Bigg]
\end{align*}

\begin{align*}
x_{I_\pm} &=
\frac{
\alpha (-1 + 2 \beta) - \lambda - 
\sqrt{\alpha^2 (1 - 2 \beta)^2 + 
2 \alpha (1 + \beta) \lambda + 
\lambda^2}
}{
\sqrt{6} \alpha \beta \lambda
}
\\[1.5em]
y_{I_\pm} &=
-\frac{1}{\sqrt{3}} \sqrt{\Big{(}
\frac{
\alpha^2 ((1 - 2 \beta)^2 + \beta (1 + \beta) \lambda^2)
+ \lambda (\lambda - 
\sqrt{
\alpha^2 (1 - 2 \beta)^2 + 2 \alpha (1 + \beta) \lambda + \lambda^2
})
}{
\alpha^2 \beta^2 \lambda^2
}
} \\
&\quad + \frac{\alpha}{\sqrt{3} \alpha^2 \beta^2 \lambda^2}
\left[
(2 - \beta) \lambda + \beta \lambda^3 +
(1 - 2 \beta) \sqrt{
\alpha^2 (1 - 2 \beta)^2 + 2 \alpha (1 + \beta) \lambda + \lambda^2
} \right. \\
&\quad\quad \left. +
\beta \lambda^2 \sqrt{
\alpha^2 (1 - 2 \beta)^2 + 2 \alpha (1 + \beta) \lambda + \lambda^2
}\Big{)}
\right]
\\[1.5em]
u_{I_\pm} &=
- \frac{1}{4 \alpha \beta \lambda}
\Bigg[
\lambda^2 + 2 \alpha \sqrt{
\alpha^2 (1 - 2 \beta)^2 + 2 \alpha (1 + \beta) \lambda + \lambda^2
}
\\
&\quad + \lambda \sqrt{
\alpha^2 (1 - 2 \beta)^2 + 2 \alpha (1 + \beta) \lambda + \lambda^2
}
+ \alpha \beta \lambda \sqrt{
\alpha^2 (1 - 2 \beta)^2 + 2 \alpha (1 + \beta) \lambda + \lambda^2
}
\\
&\quad + \alpha \lambda (3 + \beta (2 + \lambda)) + 
\alpha^2 \left(2 + \beta(-4 + \lambda + 2 \beta \lambda)\right)
\\
&\quad + \alpha \beta \lambda \sqrt{
\frac{1}{\alpha^2 \beta^2 \lambda^2}
\left(
8 (\alpha + \lambda + \alpha \beta (-2 + \lambda^2)
- \sqrt{
\alpha^2 (1 - 2 \beta)^2 + 2 \alpha (1 + \beta) \lambda + \lambda^2
})
\right.
}
\\
&\quad\quad \times (\alpha (-1 + 2 \beta) - \lambda + 
\sqrt{
\alpha^2 (1 - 2 \beta)^2 + 2 \alpha (1 + \beta) \lambda + \lambda^2
})
\\
&\quad\quad + \alpha^2 (2 + \beta(-4 + \lambda + 2 \beta \lambda)) +
\lambda (\lambda - \sqrt{
\alpha^2 (1 - 2 \beta)^2 + 2 \alpha (1 + \beta) \lambda + \lambda^2
})
\\
&\quad\quad \left. +
\alpha \left(
-2 \sqrt{
\alpha^2 (1 - 2 \beta)^2 + 2 \alpha (1 + \beta) \lambda + \lambda^2
}
+ \lambda (3 + \beta (2 + \lambda -
\sqrt{
\alpha^2 (1 - 2 \beta)^2 + 2 \alpha (1 + \beta) \lambda + \lambda^2
}))
\right)
\right)^2
\Bigg]
\end{align*}

\begin{align*}
 x_{J_\pm} &=
\frac{
\alpha (-1 + 2 \beta) - \lambda - 
\sqrt{\alpha^2 (1 - 2 \beta)^2 + 
2 \alpha (1 + \beta) \lambda + 
\lambda^2}
}{
\sqrt{6} \alpha \beta \lambda
}
\\[1.5em]
y_{J_\pm} &=
\frac{1}{\sqrt{3}} \sqrt{\Big{(}
\frac{
\alpha^2 ((1 - 2 \beta)^2 + \beta (1 + \beta) \lambda^2)
+ \lambda (\lambda - 
\sqrt{
\alpha^2 (1 - 2 \beta)^2 + 2 \alpha (1 + \beta) \lambda + \lambda^2
})
}{
\alpha^2 \beta^2 \lambda^2
}
} \\
&\quad + \frac{\alpha}{\sqrt{3} \alpha^2 \beta^2 \lambda^2}
\left[
(2 - \beta) \lambda + \beta \lambda^3 +
(1 - 2 \beta) \sqrt{
\alpha^2 (1 - 2 \beta)^2 + 2 \alpha (1 + \beta) \lambda + \lambda^2
} \right. \\
&\quad\quad \left. +
\beta \lambda^2 \sqrt{
\alpha^2 (1 - 2 \beta)^2 + 2 \alpha (1 + \beta) \lambda + \lambda^2
}\Big{)}
\right]
\\[1.5em]
u_{J_\pm} &=
- \frac{1}{4 \alpha \beta \lambda}
\Bigg[
\lambda^2 + 2 \alpha \sqrt{
\alpha^2 (1 - 2 \beta)^2 + 2 \alpha (1 + \beta) \lambda + \lambda^2
}
\\
&\quad + \lambda \sqrt{
\alpha^2 (1 - 2 \beta)^2 + 2 \alpha (1 + \beta) \lambda + \lambda^2
}
+ \alpha \beta \lambda \sqrt{
\alpha^2 (1 - 2 \beta)^2 + 2 \alpha (1 + \beta) \lambda + \lambda^2
}
\\
&\quad + \alpha \lambda (3 + \beta (2 + \lambda)) + 
\alpha^2 \left(2 + \beta(-4 + \lambda + 2 \beta \lambda)\right)
\\
&\quad + \alpha \beta \lambda \sqrt{
\frac{1}{\alpha^2 \beta^2 \lambda^2}
\left(
8 (\alpha + \lambda + \alpha \beta (-2 + \lambda^2)
- \sqrt{
\alpha^2 (1 - 2 \beta)^2 + 2 \alpha (1 + \beta) \lambda + \lambda^2
})
\right.
}
\\
&\quad\quad \times (\alpha (-1 + 2 \beta) - \lambda + 
\sqrt{
\alpha^2 (1 - 2 \beta)^2 + 2 \alpha (1 + \beta) \lambda + \lambda^2
})
\\
&\quad\quad + \alpha^2 (2 + \beta(-4 + \lambda + 2 \beta \lambda)) +
\lambda (\lambda - \sqrt{
\alpha^2 (1 - 2 \beta)^2 + 2 \alpha (1 + \beta) \lambda + \lambda^2
})
\\
&\quad\quad \left. +
\alpha \left(
-2 \sqrt{
\alpha^2 (1 - 2 \beta)^2 + 2 \alpha (1 + \beta) \lambda + \lambda^2
}
+ \lambda (3 + \beta (2 + \lambda -
\sqrt{
\alpha^2 (1 - 2 \beta)^2 + 2 \alpha (1 + \beta) \lambda + \lambda^2
}))
\right)
\right)^2
\Bigg]
\end{align*}
}

\begin{table}[H]
\centering
\begin{tabular}{|c|c|c|}
\hline
\textbf{Critical } & \textbf{Critical matter-density}  & \textbf{Total EoS parameter} \\
\textbf{ Points} & ($\boldsymbol{\Omega_{\rm m}}$)  &  ($\boldsymbol{\omega_{\rm tot}}$) \\
\hline
$A_{\pm}$ & $0$ & $1$ \\
$B_{\pm}$ & $0$ & $1$ \\
$C_{\pm}$ 
& $1 - \dfrac{\left(1 + \sqrt{1 + 2 \alpha^2 (-2 + \beta) \beta}\right)^2}{6 \alpha^2 \beta^2}$ 
& $\dfrac{\left(1 + \sqrt{1 + 2 \alpha^2 (-2 + \beta) \beta}\right)^2}{6 \alpha^2 \beta^2}$ \\
$D_{\pm}$ 
& $1 - \dfrac{\left(-1 + \sqrt{1 + 2 \alpha^2 (-2 + \beta) \beta}\right)^2}{6 \alpha^2 \beta^2}$ 
& $\dfrac{\left(-1 + \sqrt{1 + 2 \alpha^2 (-2 + \beta) \beta}\right)^2}{6 \alpha^2 \beta^2}$ \\
$E_{\pm}$ 
& $0$ 
& $\frac{\lambda^2-3}{3}$ \\
$F_{\pm}$ 
& $0$ 
& $\frac{\lambda^2-3}{3}$ \\
$G_{\pm}$ & $\Omega_m (G_{\pm})$ & $\omega_{\rm tot} (G_{\pm})$ \\
$H_{\pm}$ & $\Omega_m (H_{\pm})$ & $\omega_{\rm tot} (H_{\pm})$ \\
$I_{\pm}$ & $\Omega_m (I_{\pm})$ & $\omega_{\rm tot} (I_{\pm})$ \\
$J_{\pm}$ & $\Omega_m (J_{\pm})$ & $\omega_{\rm tot} (J_{\pm})$ \\
\hline
\end{tabular}
\caption{Matter density parameter \( \Omega_m \) and total equation of state \( \omega_{\text{tot}} \) evaluated at the critical points (CPs).}
\label{tab:2}
\end{table}
 
\textbf{\underline{Explicit forms for $\Omega_m$ and $\omega_{\rm tot} $ at the critical points:}}
{\scriptsize
\begin{align*}
\Omega_{m}(G_\pm) = \Omega_{m}(H_\pm) =
\frac{1}{3 \alpha^2 \beta^2 \lambda^2} \bigg(&
\alpha^2 (-1 + 2 \beta)\left(2 + \beta(-4 + \lambda^2)\right) \\
& + 2 \lambda \left( -\lambda + 
\sqrt{ \alpha^2 (1 - 2 \beta)^2 + 2 \alpha (1 + \beta)\lambda + \lambda^2 } \right) \\
& + \alpha \Big(
2 (-2 + \beta)\lambda 
- \beta \lambda^3 
+ 2 (1 - 2 \beta) 
\sqrt{ \alpha^2 (1 - 2 \beta)^2 + 2 \alpha (1 + \beta)\lambda + \lambda^2 } \\
& \quad + \beta \lambda^2 
\sqrt{ \alpha^2 (1 - 2 \beta)^2 + 2 \alpha (1 + \beta)\lambda + \lambda^2 }
\Big)
\bigg)
\end{align*}

\begin{align*}
\omega_{\rm tot}(G_\pm) = \omega_{\rm tot}(H_\pm) =
- \frac{1}{3 \alpha \beta} \bigg(
\alpha + \alpha \beta + \lambda 
- \sqrt{ 
\alpha^2 (1 - 2 \beta)^2 
+ 2 \alpha (1 + \beta)\lambda 
+ \lambda^2 
}
\bigg)
\end{align*}

\begin{align*}
\Omega_{m}(I_\pm) = \Omega_{m}(J_\pm) =
\frac{1}{3 \alpha^2 \beta^2 \lambda^2} \bigg(&
\alpha^2 (-1 + 2 \beta)\left(2 + \beta(-4 + \lambda^2)\right) \\
& - 2 \lambda \left( \lambda + 
\sqrt{ \alpha^2 (1 - 2 \beta)^2 + 2 \alpha (1 + \beta)\lambda + \lambda^2 } \right) \\
& - \alpha \Big(
-2 (-2 + \beta)\lambda 
+ \beta \lambda^3 
+ 2 (1 - 2 \beta) 
\sqrt{ \alpha^2 (1 - 2 \beta)^2 + 2 \alpha (1 + \beta)\lambda + \lambda^2 } \\
& \quad + \beta \lambda^2 
\sqrt{ \alpha^2 (1 - 2 \beta)^2 + 2 \alpha (1 + \beta)\lambda + \lambda^2 }
\Big)
\bigg)
\end{align*}

\begin{align*}
\omega_{\rm tot}(I_\pm) = \omega_{\rm tot}(J_\pm) = 
- \frac{1}{3 \alpha \beta} \bigg(
\alpha + \alpha \beta + \lambda 
+ \sqrt{ 
\alpha^2 (1 - 2 \beta)^2 
+ 2 \alpha (1 + \beta)\lambda 
+ \lambda^2 
}
\bigg)
\end{align*}
}

\section{Results and Discussion}
\label{sec:ROD}
A key feature of the curvature-modulated interaction model lies in the role of the parameter \( \beta \), which appears in the coupling term \( Q_0 = \alpha \kappa \rho_m \dot{\phi} \left(1 - \frac{\beta}{6H^2} R\right) \). This parameter governs how strongly the interaction between dark energy and dark matter is influenced by the spacetime curvature. When \( \beta = 0 \), the coupling depends solely on the scalar field’s time evolution and dark matter density, with no geometric feedback. However, for non-zero \( \beta \), the Ricci scalar \( R \) contributes dynamically, making the interaction sensitive to the background geometry of the universe. To illustrate the impact of this modulation, we study the evolution of the dimensionless quantity \( \gamma R = \beta R / 6H^2 \), which encapsulates the effective strength of curvature regulation. In scenarios with small \( \beta \), such as \( \beta = 0.05 \), this term remains well below unity across cosmic history, implying a weakly coupled regime akin to minimal coupling. In contrast, a moderate value like \( \beta = 0.5 \) leads to \( \gamma R \sim 1 \) near \( a \sim 1 \), indicating that curvature feedback becomes dynamically significant just as the universe transitions into accelerated expansion. Based on this schematic behavior, we adopt \( \beta = 0.05 \) and \( \beta = 0.5 \) as benchmark values to systematically compare models with negligible versus substantial curvature modulation. This comparative framework enables us to understand how the evolving geometry of the universe can activate or suppress the energy transfer in the dark sector and shape both the background evolution and structure formation.

\subsection{Stability analysis for critical points along with their cosmological properties}

The full autonomous dynamical system and the associated critical point structure are governed by three free parameters: the interaction strength \( \alpha \), the potential slope \( \lambda \), and the curvature modulation parameter \( \beta \). The expressions for the coordinates of the critical points \( (x, y, u) \), as well as the derived quantities such as \( \Omega_m \) and \( \omega_{\rm tot} \), are all explicitly dependent on these three parameters. However, simultaneous analysis in the full three-dimensional \( (\alpha, \lambda, \beta) \)-space can be analytically and numerically intractable due to the nonlinear structure of the system and the complexity of the stability conditions. To address this, a more tractable yet informative approach is adopted by fixing the value of one of the parameters—specifically, the curvature modulation parameter \( \beta \)—and performing a systematic analysis in the remaining \( (\alpha, \lambda) \)-plane. This choice is physically motivated: \( \beta \) directly governs the role of the Ricci scalar \( R \) in modulating the interaction and determines the degree of curvature feedback in the system. By selecting representative values of \( \beta \), such as \( \beta = 0.05 \) for weak modulation and \( \beta = 0.5 \) for moderate curvature sensitivity, we can isolate and explore how the coupled dynamics and critical point structure respond to varying field strength \( \alpha \) and potential steepness \( \lambda \), while maintaining a controlled geometric background. This reduced-parameter analysis enables clear visualization of phase-space trajectories, stability conditions, and cosmological viability across meaningful subspaces of the theory. A detailed analysis of the critical points, including their stability properties and key cosmological characteristics, is provided below.\\

\begin{itemize}
    \item \textbf{Points \( A_\pm \):} These critical points are always real and exist and represent purely kinetic scalar field-dominated solutions, with \( \Omega_m = 0 \) and \( \omega_{\rm tot} = 1 \), indicating a stiff fluid–like equation of state. They exist independently of the model parameters \( \alpha \), \( \lambda \), and \( \beta \), and are associated with a no-growth regime since the perturbation growth rate remains fixed at \( u = 0 \). While dynamically significant as early-time attractors or saddle points, they are dynamically disfavored due to their incompatibility with a matter-dominated epoch. These points are not capable of yielding stable late-time solutions.

    \item \textbf{Points \( B_\pm \):} These critical points are always real and exist for all real $\alpha, \beta$ and also fully scalar field-dominated with \( \Omega_m = 0 \) and \( y = 0 \), but unlike \( A_\pm \), they allow a non-zero perturbation growth rate \( u = 1 + \alpha(\pm \sqrt{6} + \beta) \), which may indicate growing or decaying modes depending on the coupling strength \( \alpha \). Despite sharing the same stiff equation of state \( \omega_{\rm tot} = 1 \), these points are unstable in nature. However, even within favorable ranges, they fail to support stable accelerated evolution and typically manifest as unstable type points in the phase space.

    \item \textbf{Points \( C_\pm \):} Existence criteria for these critical points at values of \( \alpha \), \( \beta \) and \( \lambda \) satisfying\\
    $ \alpha^2 (2\beta - 1)\left[\beta(\lambda^2 - 4) + 2\right] \geq 2\lambda \left( \lambda + \sqrt{ \alpha^2 (2\beta - 1)^2 + 2\alpha\lambda(\beta + 1) + \lambda^2 } \right)$ and $1 + 2\alpha^2(\beta - 2)\beta \geq 0$.   These critical points exhibit more intricate dependence on the curvature-modulated coupling, with both the coordinates \( x_{C_\pm} \) and the perturbation growth rate \( u_{C_\pm} \) depending nonlinearly on \( \alpha \), \( \beta \), and their combinations. They correspond to scenarios where the potential energy vanishes, but the kinetic energy is such that \( \Omega_m \neq 0 \), allowing for partial matter–field scaling behavior. The equation of state \( \omega_{\rm tot} \) also reflects this mixed composition and can deviate from unity. Despite their richness, these points do not correspond to attractor solutions and remain dynamically unstable.

    \item \textbf{Points \( D_\pm \):}  Existence criteria for these critical points at values of \( \alpha \), \( \beta \) and \( \lambda \) satisfying \\
$\alpha^2 (2\beta - 1)\left[\beta(\lambda^2 - 4) + 2\right] 
\geq 2\lambda \left( \lambda - \sqrt{ \alpha^2 (2\beta - 1)^2 + 2\alpha\lambda(\beta + 1) + \lambda^2 } \right)$ and $1 + 2\alpha^2(\beta - 2)\beta \geq 0$.   Similar in structure to \( C_\pm \), these points also involve vanishing potential energy and retain non-trivial \( x \) and \( u \) components. However, the square root expressions in their definitions introduce branches in their dynamics, which can distinguish between growing and decaying modes. The matter density \( \Omega_m \) is typically non-zero and decreases with increasing coupling, indicating potential relevance in mediating transitions between kinetic and potential dominated epochs. These points may act as transient saddle configurations but do not support stable, late-time acceleration and thus are not viable as cosmological attractors.
    
    \item \textbf{Points \( E_\pm \):} Existence criteria for these critical points are $\lambda^2 \leq 6$. These critical points represent purely scalar field–dominated configurations, defined by the coordinates \( x = \lambda / \sqrt{6} \), \( y = \pm \sqrt{1 - \lambda^2 / 6} \), and \( u = 0 \). They correspond to a phase where \( \Omega_m = 0 \) and the scalar field fully drives the cosmic expansion. The total equation of state is \( \omega_{\rm tot} = \lambda^2/3 - 1 \), which leads to accelerated expansion for \( \lambda^2 < 2 \). Since the perturbation growth rate \( u \) vanishes, these points imply a frozen matter perturbation sector, making them ideal candidates for the late-time accelerated attractor in scenarios where structure formation is effectively halted. Fig.~\ref{fig:1} displays the allowed parameter regions for these points in the \( \alpha\text{--}\lambda \) plane, for both \( \beta = 0.05 \) (left panel) and \( \beta = 0.5 \) (right panel). The allowed regions are broad and continuous, particularly for small \( |\lambda| \), confirming the viability of \( E_\pm \) across a wide range of interaction strengths and curvature modulation. The independence of \( u \) from \( \alpha \) and \( \beta \) further highlights that these solutions are purely background-driven and geometry-insensitive.

\begin{figure}[H]
    \centering
        \includegraphics[width=0.8\textwidth]{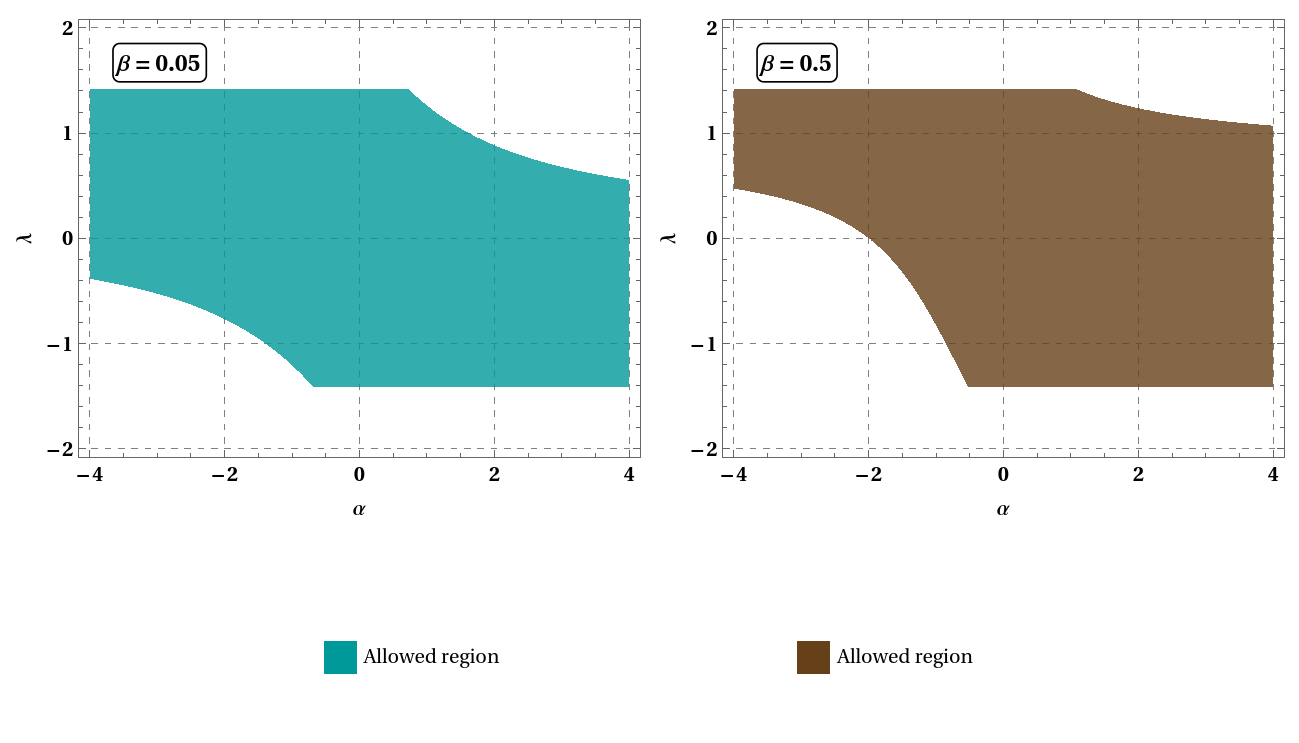}
        \caption{Allowed regions in the \( \alpha\text{--}\lambda \) parameter space for benchmark values of \( \beta \). 
The shaded regions satisfy all of the following: stability conditions, \( 0 < \Omega_m < 1 \) and \( -1 < \omega_{\text{tot}} < -1/3 \). 
Left and right panels correspond to \( \beta = 0.05 \) and \( \beta = 0.5 \), respectively. }
        \label{fig:1}
    \end{figure}

    \item \textbf{Points \( F_\pm \):} Existence criteria for these critical points are $\lambda^2 \leq 6$. These points also satisfy \( x = \lambda / \sqrt{6} \) and \( y = \pm \sqrt{1 - \lambda^2 / 6} \), similar to \( E_\pm \), but differ crucially in that they allow a non-zero perturbation growth rate $
    u = \frac{1}{2} \left( -4 + 2\alpha \lambda + \lambda^2 + \alpha \beta (-4 + \lambda^2) \right)$,
    which depends on all three model parameters. These configurations maintain \( \Omega_m = 0 \), but enable curvature-sensitive structure growth even during scalar field domination. Figure~\ref{fig:2} shows the allowed regions in the \( \alpha\text{--}\lambda \) plane where \( F_\pm \) exist and yield physically meaningful \( y \in \mathbb{R} \) and \( \omega_{\rm tot} < -1/3 \). Importantly, the plots are shaded according to the sign of \( u \): regions with \( u > 0 \) (light blue) correspond to scenarios where growth persists, whereas \( u < 0 \) (light red) indicates suppressed growth. The black dashed mark where \( u = 0 \), i.e., the boundary with \( E_\pm \). For both \( \beta = 0.05 \) and \( \beta = 0.5 \), the \( F_\pm \) points occupy distinct, bounded regions in parameter space, and their viability is highly sensitive to both \( \alpha \) and \( \beta \). This makes \( F_\pm \) particularly interesting for models that aim to interpolate between growth-compatible and growth-suppressed cosmologies through the modulation of curvature. Their ability to support non-zero \( u \) during acceleration provides a richer phenomenology than \( E_\pm \), and they are especially relevant for addressing the transition between matter-dominated growth and dark energy–driven suppression of clustering.
    
     \begin{figure}[H]
    \centering
        \includegraphics[width=0.8\textwidth]{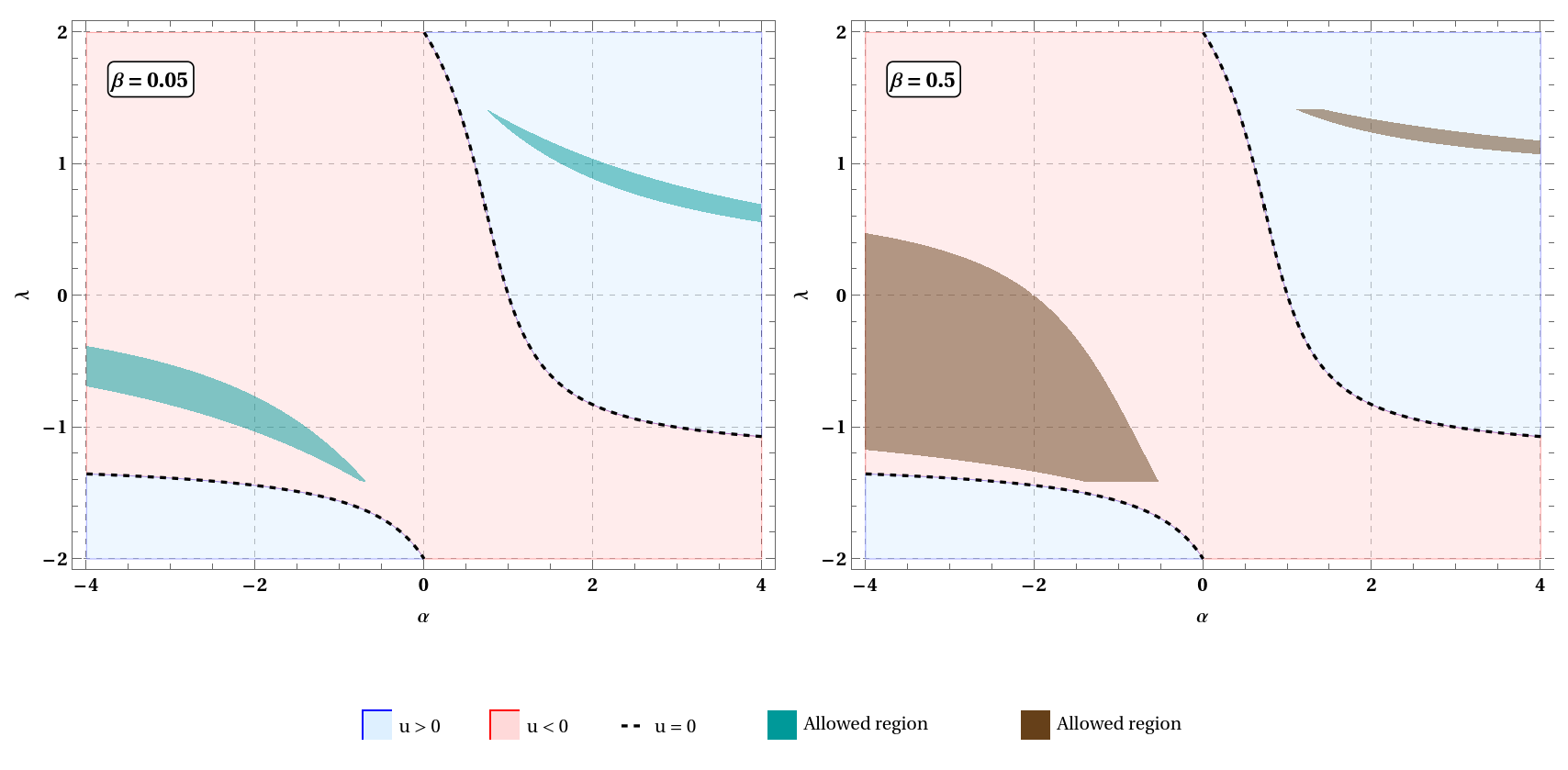}
        \caption{Allowed regions in the \( \alpha\text{--}\lambda \) parameter space for benchmark values of \( \beta \). 
The shaded regions satisfy all of the following: stability conditions, \( 0 < \Omega_m < 1 \) and \( -1 < \omega_{\text{tot}} < -1/3 \). 
Left and right panels correspond to \( \beta = 0.05 \) and \( \beta = 0.5 \), respectively. 
Colored overlays indicate different signs of the growth rate \( u \) for the \( F_\pm \) points, with black dashed  marking \( u = 0 \).} 
        \label{fig:2}
    \end{figure}

  \item \textbf{Points \( G_\pm \):} Existence criteria for these critical points are $\alpha^2 (2\beta - 1)^2 + 2\alpha\lambda(\beta + 1) + \lambda^2 \geq 0$.  These critical points correspond to mixed dark energy–dark matter scaling solutions characterized by non-zero values of all three dynamical variables \( x \), \( y \), and \( u \), with highly non-linear dependence on the model parameters \( \alpha \), \( \lambda \), and \( \beta \). The matter density parameter \( \Omega_m(G_\pm) \) and total equation of state \( \omega_{\rm tot}(G_\pm) \) take compact analytical forms, while the explicit coordinates \( x_{G_\pm} \), \( y_{G_\pm} \), and \( u_{G_\pm} \) are algebraically complex and include nested square roots. Notably, \( \omega_{\rm tot}(G_\pm) \) is always negative and depends on the interplay of interaction strength and curvature modulation as, $\omega_{\rm tot}(G_\pm) = -\frac{1}{3 \alpha \beta} \left( \alpha + \alpha \beta + \lambda - \sqrt{\alpha^2 (1 - 2\beta)^2 + 2\alpha (1 + \beta) \lambda + \lambda^2} \right)$. These points allow for non-zero \( \Omega_m \) and thus may describe intermediate phases in cosmic evolution where energy densities of dark energy and matter scale comparably. Despite their complex structure, they are typically found to be saddle points in the dynamical system and do not serve as late-time attractors. Their existence provides dynamical bridges between early-time and accelerating phases, and they may influence the trajectory paths in phase space through transient matter-dominated behavior or delayed acceleration onset. The perturbation growth rate \( u_{G_\pm} \) is generically non-zero and parameter-sensitive, indicating that these points could permit structure formation under specific conditions. However, due to the lack of asymptotic stability, \( G_\pm \) are not viable endpoints of cosmic evolution but contribute to the richness of the intermediate dynamical landscape.

\item \textbf{Points \( H_\pm \):} Existence criteria for these critical points are $\alpha^2 (2\beta - 1)^2 + 2\alpha\lambda(\beta + 1) + \lambda^2 \geq 0$. These critical points share many structural similarities with the \( G_\pm \) points, including the same total matter density \( \Omega_m(H_\pm) \) and total equation of state \( \omega_{\rm tot}(H_\pm) \). However, their explicit coordinate expressions—particularly for \( x_{H_\pm} \), \( y_{H_\pm} \), and \( u_{H_\pm} \)—are distinct and involve nested parameter-dependent square roots. The perturbation growth rate \( u_{H_\pm} \) is significantly more intricate than that of \( G_\pm \), reflecting the rich internal coupling between the scalar field, background curvature, and dark matter. Although these points admit non-zero \( \Omega_m \), the scaling behavior does not typically lead to stable cosmic evolution. Rather, \( H_\pm \) function as intermediate saddle points, dynamically bridging earlier or transient matter-dominated regimes with later dark energy–dominated attractors. The non-vanishing and parameter-sensitive value of \( u_{H_\pm} \) suggests that structure formation is still active around these points, depending on the sign and magnitude of \( u \). However, their lack of asymptotic stability excludes them as viable late-time attractors. Overall, the \( H_\pm \) points enrich the phase space structure by offering a non-trivial yet transient regime where all cosmic components contribute meaningfully.

    \item \textbf{Points \( I_\pm \):} Existence criteria for these critical points are $\alpha^2 (2\beta - 1)^2 + 2\alpha\lambda(\beta + 1) + \lambda^2 \geq 0$. These critical points form a counterpart to the \( H_\pm \) solutions, but with qualitatively distinct dynamical characteristics. They are described by highly non-linear expressions for \( x_{I_\pm} \), \( y_{I_\pm} \), and \( u_{I_\pm} \), with the growth rate \( u \) being sensitive to all three model parameters \( \alpha \), \( \lambda \), and \( \beta \). The matter density parameter \( \Omega_m(I_\pm) \) and equation of state \( \omega_{\rm tot}(I_\pm) \) are analytically compact and match those of the \( J_\pm \) points:
    $\omega_{\rm tot}(I_\pm) = -\frac{1}{3\alpha\beta} \left( \alpha + \alpha\beta + \lambda + \sqrt{\alpha^2(1 - 2\beta)^2 + 2\alpha(1 + \beta)\lambda + \lambda^2} \right),$
    indicating a negative but non-accelerating equation of state for most parameter choices. Unlike scaling attractors or saddle points, the \( I_\pm \) points exhibit unstable behavior in most regions of the parameter space, leading to divergent trajectories in their vicinity. The growth rate \( u_{I_\pm} \) is large and generally positive, which implies strong structure formation, but in a way that is not balanced by a stable dynamical configuration. As such, these points are not viable as late-time attractors or scaling solutions. Their instability makes them dynamically short-lived, and they tend to serve as repeller points that redirect trajectories toward more stable regions such as \( E_\pm \) or curvature-sensitive saddle points like \( G_\pm \) and \( H_\pm \). Although not physically favorable in the cosmological sense, the presence of \( I_\pm \) points adds dynamical richness to the overall phase space and offers insights into the behavior of the system under high-curvature or strongly coupled limits.

    \item \textbf{Points \( J_\pm \):} Existence criteria for these critical points are $\alpha^2 (2\beta - 1)^2 + 2\alpha\lambda(\beta + 1) + \lambda^2 \geq 0$. These critical points closely mirror the structure of \( I_\pm \), with identical forms for \( \Omega_m \) and \( \omega_{\rm tot} \), but differ significantly in dynamical behavior. Their coordinates—\( x_{J_\pm} \), \( y_{J_\pm} \), and \( u_{J_\pm} \)—are highly non-linear and depend on all three model parameters \( \alpha \), \( \lambda \), and \( \beta \), with \( u_{J_\pm} \) containing intricate combinations of square roots and cross terms. The total equation of state remains negative, $ \omega_{\rm tot}(J_\pm) = -\frac{1}{3\alpha\beta} \left( \alpha + \alpha\beta + \lambda + \sqrt{\alpha^2(1 - 2\beta)^2 + 2\alpha(1 + \beta)\lambda + \lambda^2} \right),$ while the growth rate \( u_{J_\pm} \) is generically negative for physically allowed parameter choices. As shown in Fig.~\ref{fig:3}, the allowed and stable regions in the \( \alpha\text{--}\lambda \) parameter space are clearly visible for both \( \beta = 0.05 \) and \( \beta = 0.5 \). In contrast to their unstable counterparts \( I_\pm \), the \( J_\pm \) points occupy compact but well-defined regions where all conditions for physical viability—namely \( 0 < \Omega_m < 1 \), \( -1 < \omega_{\rm tot} < -1/3 \), and \( u > 0 \)—are satisfied, and additionally, the eigenvalue analysis confirms their stable node or spiral behavior. These points thus represent viable late-time attractor solutions in the presence of non-zero curvature modulation. The transition from instability in \( I_\pm \) to stability in \( J_\pm \), visible in the shaded allowed zones of the stability plots, highlights the crucial role of the curvature modulation parameter \( \beta \) in screening and dynamically regulating the coupling. From a cosmological perspective, \( J_\pm \) are among the most important critical points in the model, enabling both late-time acceleration and suppression of structure formation in a dynamically consistent and allowed manner.

\begin{figure}[H]
    \centering
        \centering
        \includegraphics[width=0.8\textwidth]{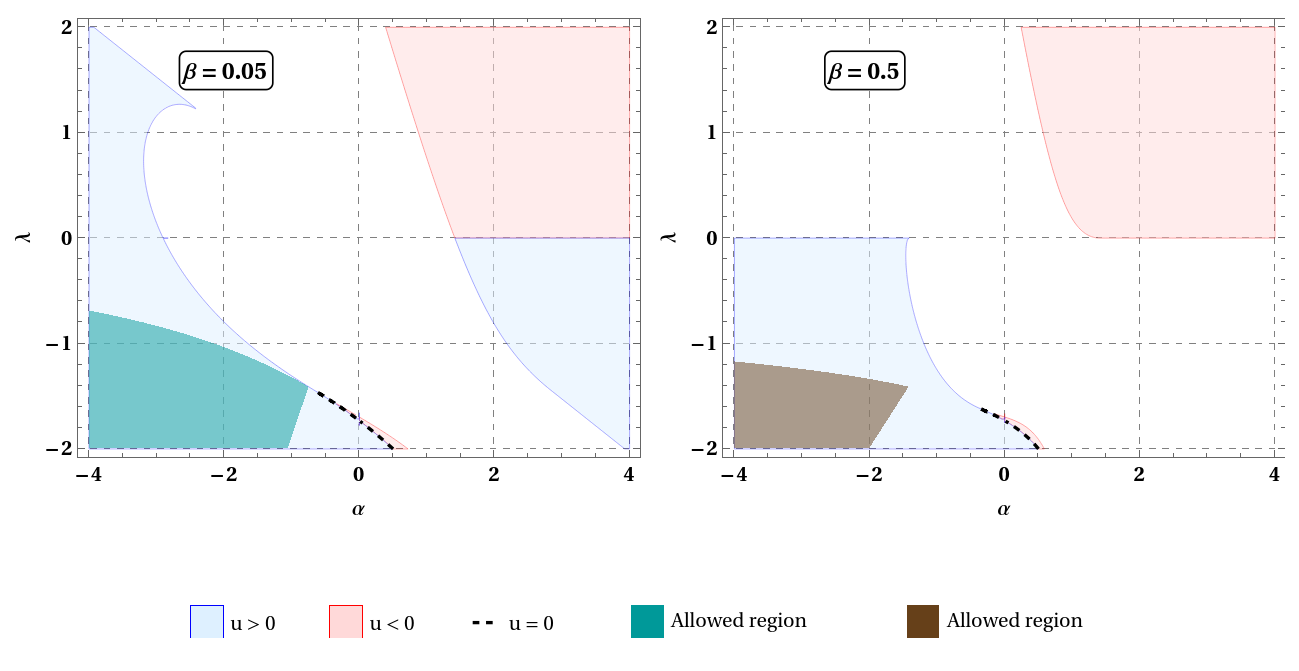}
        \caption{Allowed regions in the \( \alpha\text{--}\lambda \) parameter space for benchmark values of \( \beta \). 
The shaded regions satisfy all of the following: stability conditions, \( 0 < \Omega_m < 1 \) and \( -1 < \omega_{\text{tot}} < -1/3 \). 
Left and right panels correspond to \( \beta = 0.05 \) and \( \beta = 0.5 \), respectively. 
Colored overlays indicate different signs of the growth rate \( u \) for the \( J_\pm \) points, with black dashed marking \( u = 0 \).}
        \label{fig:3}

\end{figure}
\end{itemize}

\subsection{Choice of model parameters ($ \beta, \alpha, \lambda$)}
\begin{figure}[H]
    \centering
    \includegraphics[width=0.9\textwidth]{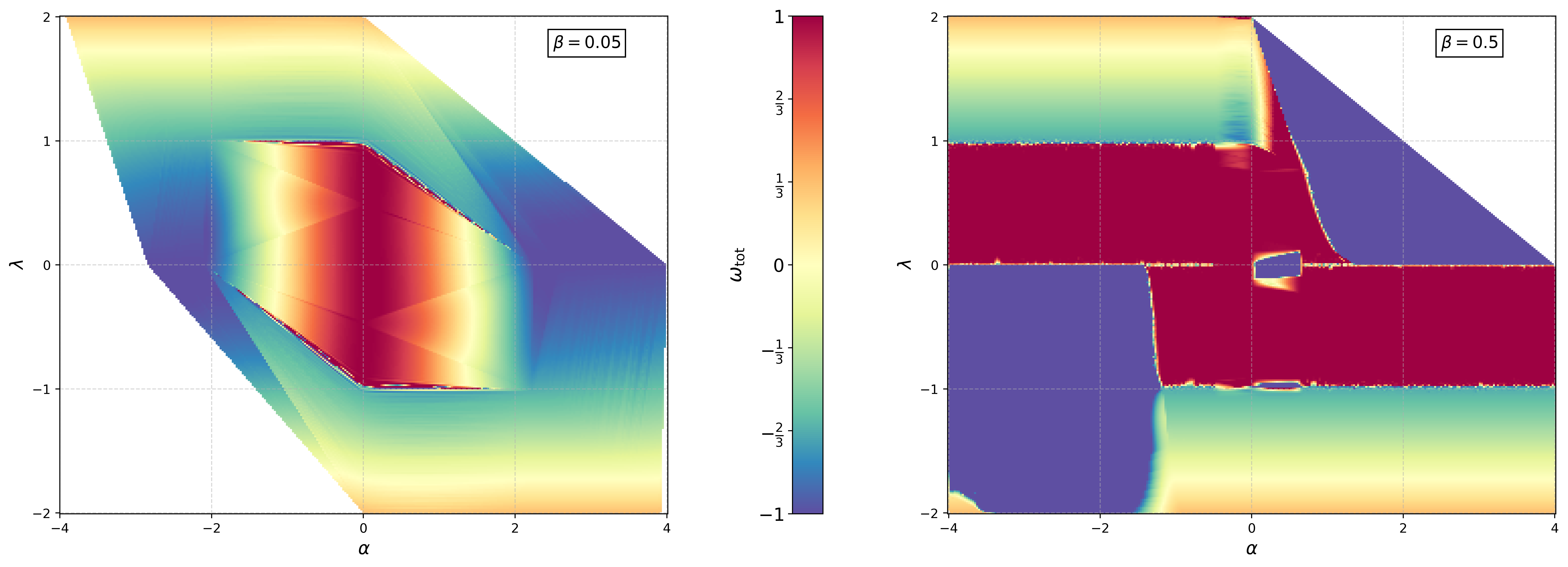}
    \caption{Density plots showing the variation of the grand equation of state parameter \( \omega_{\text{tot}} \) 
in the \( \alpha-\lambda \) parameter space. 
Left and right panels correspond to \( \beta = 0.05 \) and \( \beta = 0.5 \), respectively.}
    \label{fig:4}
\end{figure}

To investigate the cosmological implications of our interacting dark energy–dark matter framework, we explore the $(\alpha, \lambda)$ parameter space for fixed values of the curvature-modulating parameter $\beta$, with the total effective equation of state $\omega_{\text{tot}}$ used as a probe for the late-time dynamics. The interaction is encoded in the form $Q_0 = \alpha \kappa \rho_m \dot{\phi} \left(1 - \frac{\beta R}{6H^2} \right)$, where $\alpha$ controls the overall coupling strength, while $\beta$ determines the sensitivity of the interaction to the spacetime curvature through the Ricci scalar $R$. Importantly, even when $\beta = 0$, the interaction remains non-minimal due to the $\rho_m \dot{\phi}$ dependence, implying a direct exchange of energy-momentum between dark matter and the scalar field. In this work, we consider two physically motivated benchmarks: (i) $\beta = 0.05$, representing a weakly curvature-modulated interaction where the primary exchange is governed by the scalar field derivative and dark matter density, and (ii) $\beta = 0.5$, which significantly enhances curvature dependence and allows for stronger background sensitivity. For each benchmark, we perform a comprehensive scan over the $(\alpha, \lambda)$ parameter space and compute $\omega_{\text{tot}}$ as a function of these parameters. The resulting parameter-space diagrams, shown in fig.~\ref{fig:4}, reveal the regions that permit a late-time accelerating universe (i.e., $\omega_{\text{tot}} < -1/3$). For $\beta = 0.05$, a narrow, symmetric region supports acceleration, indicating that even weak curvature modulation restricts the viable range. For $\beta = 0.5$, the landscape is more fragmented and complex, featuring multiple viable pockets and sharp transitions. From these results, we extract benchmark points that satisfy three key criteria: (a) the presence of a stable critical point, (b) critical matter density in between 0 to 1, and (c) compatibility with accelerated cosmic expansion. These benchmarks form the basis of the detailed dynamical and observational analysis developed in the following sections.

\begin{figure}[H]
    \centering
        \includegraphics[width=0.6\textwidth]{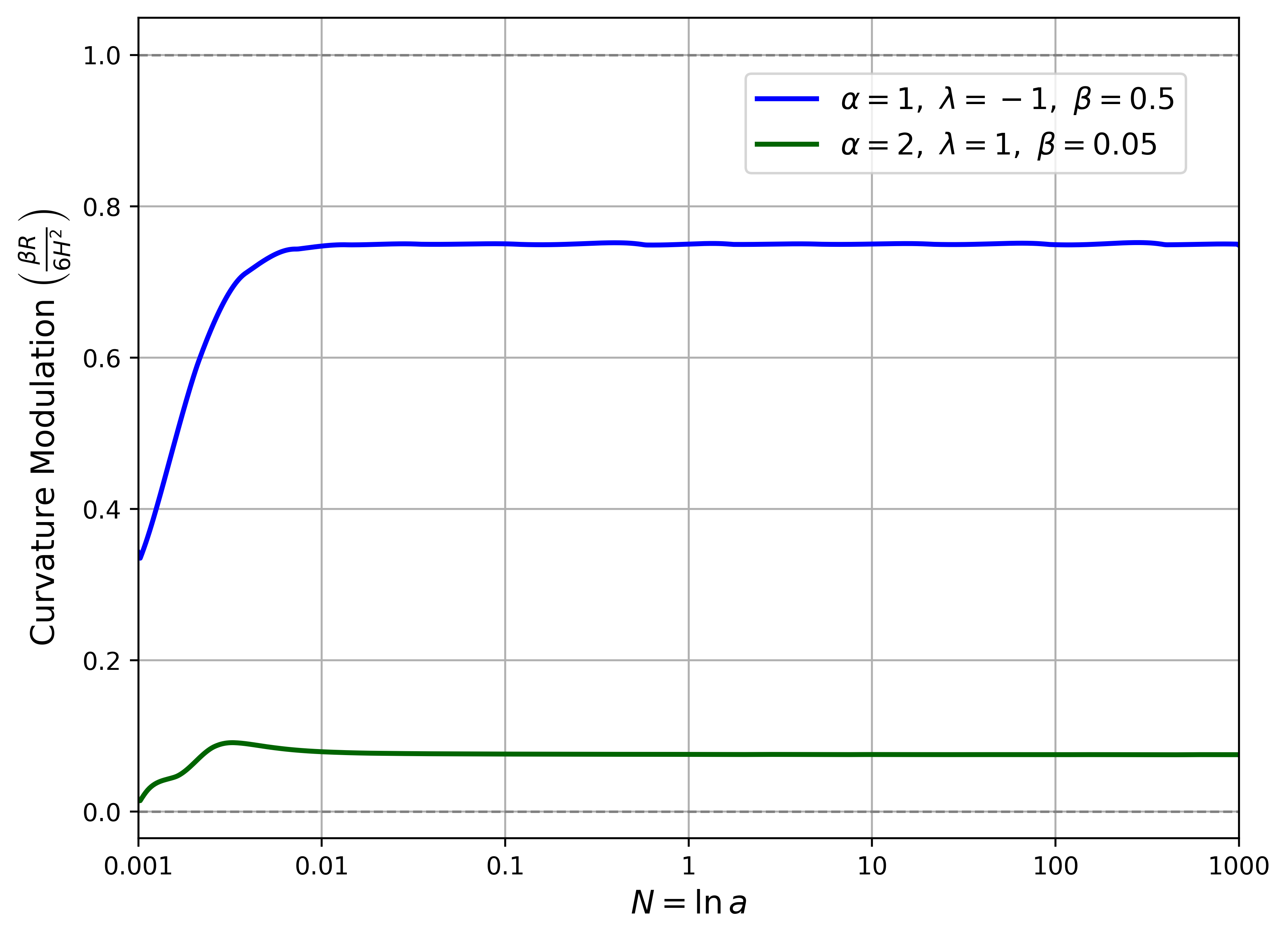}
       \caption{Variation of curvature modulation parameter with e-folding number. Curves correspond to $(\alpha = 1,\ \lambda = -1,\ \beta = 0.5)$ (blue) and $(\alpha = 2,\ \lambda = 1,\ \beta = 0.05)$ (green).
}
        \label{fig:beta}
   \end{figure}

To gain deeper insight into the curvature-regulated dynamics of the interaction term, we examine two representative benchmark scenarios: $(\alpha = 1, \lambda = -1, \beta = 0.5)$ and $(\alpha = 2, \lambda = 1, \beta = 0.05)$, chosen to illustrate the influence of curvature sensitivity encoded through the modulation parameter $\frac{\beta R}{6H^2}$. As shown in fig. \ref{fig:beta}, for $\beta = 0.5$, the curvature contribution becomes significant at early times and saturates around 0.7, indicating a strong coupling between the scalar field evolution and the background geometry, especially during late-time dynamics. This enhances the model’s sensitivity to curvature, allowing for rich dynamical behavior as reflected in the complex $(\alpha, \lambda)$ landscape. In contrast, the case with $\beta = 0.05$ exhibits negligible modulation throughout cosmic evolution, suggesting that the interaction is effectively governed by the scalar field derivative $\dot{\phi}$ and dark matter density $\rho_m$, with minimal curvature feedback. These behaviors emphasize an important theoretical constraint: to ensure physical consistency and avoid runaway effects from excessive curvature dependence, the modulation parameter $\beta$ must remain below unity. Restricting $\beta$ to small or moderate values not only preserves stability but also allows the curvature-modulated interaction to remain a well-behaved extension of standard coupled dark energy models.

\subsection{Phase space trajectories and background evolution }
To probe the dynamical and observational viability of interacting dark energy models with curvature-modulated couplings, we investigate two distinct benchmark scenarios characterized by different combinations of interaction strength \( \alpha \), potential slope \( \lambda \), and curvature modulation parameter \( \beta \). These benchmark choices are not arbitrary—they are strategically selected to capture the contrasting dynamical behaviors enabled by the interplay between the scalar field and the evolving geometry of spacetime. A unique feature of our approach lies in the construction and analysis of a three-dimensional phase space spanned by the variables \( (x, y, u) \), where \( x \) encodes the normalized derivative of the scalar field and quantifies its kinetic contribution, \( y \) measures the normalized potential energy contribution of the scalar field, and \( u \) represents the logarithmic growth rate of matter perturbations. This triad captures a unified picture of the background evolution (via \( x \) and \( y \)) and the linear structure formation (via \( u \)), offering a deeper and more holistic understanding of the coupled dark sector dynamics. The 3D dynamical system exhibits rich behavior across these axes, where late-time attractors and repellers govern the fate of the universe across different cosmological epochs. Due to the symmetry of the autonomous equations under \( y \rightarrow -y \), we restrict the analysis to the semi-plane \( y \geq 0 \). Furthermore, the physically allowed region is bounded by the constraint \( x^2 + y^2 \leq 1 \), which follows from requiring \( 0 < \Omega_{\rm m} < 1 \), ensuring consistency with realistic cosmological conditions. Thus, the relevant phase-space domain is confined to the set \( \{(x, y, u) \in \mathbb{R}^3 \mid -1 \leq x \leq 1, 0 \leq y \leq 1, x^2 + y^2 \leq 1\} \). While \( x \) and \( y \) are bounded by construction, the growth rate variable \( u \) can, in principle, diverge. To address this, we performed a compactification transformation \( u \rightarrow U = \tan^{-1}u \), mapping the unbounded real line to a finite interval \( U \in (-\frac{\pi}{2}, \frac{\pi}{2}) \). Through this transformation, we verified the absence of critical points at infinity, confirming that the dynamical behavior remains well-defined within the finite compactified domain. This comprehensive three-dimensional phase-space framework thus provides a global characterization of the system, capturing its full dynamical structure. Furthermore, it brings to light the pivotal role of scalar field dynamics in shaping the evolution of structure, enabling us to trace the cosmological transition from early kinetic or matter-dominated epochs to the eventual scalar-field-dominated accelerated phase. The resulting picture offers an internally consistent and geometrically complete description of cosmic evolution under curvature-sensitive interactions.\\

To complement the phase-space analysis and provide a broader view of model viability, fig.~\ref{fig:4} presents density plots of the total equation of state parameter \( \omega_{\rm tot} \) in the \( \alpha\text{--}\lambda \) parameter space for two fixed values of the curvature modulation parameter \( \beta \), with the left and right panels corresponding to \( \beta = 0.05 \) and \( \beta = 0.5 \), respectively. These plots reveal distinct regions associated with accelerated expansion, scaling behavior, and stiff regimes, and help identify dynamically favorable combinations of model parameters. Based on these global features, two representative benchmark points were selected for detailed exploration in the subsequent sections: \( (\alpha = 2, \lambda = 1, \beta = 0.05) \), which features strong coupling and mild curvature suppression, and \( (\alpha = 1, \lambda = -1, \beta = 0.5) \), which embodies moderate coupling and strong curvature screening. These benchmarks lie within phenomenologically viable zones and span contrasting dynamical behaviors, enabling a comprehensive assessment of how interaction strength and curvature modulation jointly influence the background dynamics, structure growth, and late-time acceleration.

\subsubsection*{\textbf{Case I:} \boldmath$\boldsymbol{\alpha = 1,\ \lambda = -1,\ \beta = 0.5}$}

\begin{figure}[H]
\centerline{$\begin{array}{cc}
\includegraphics[width=0.465\textwidth]{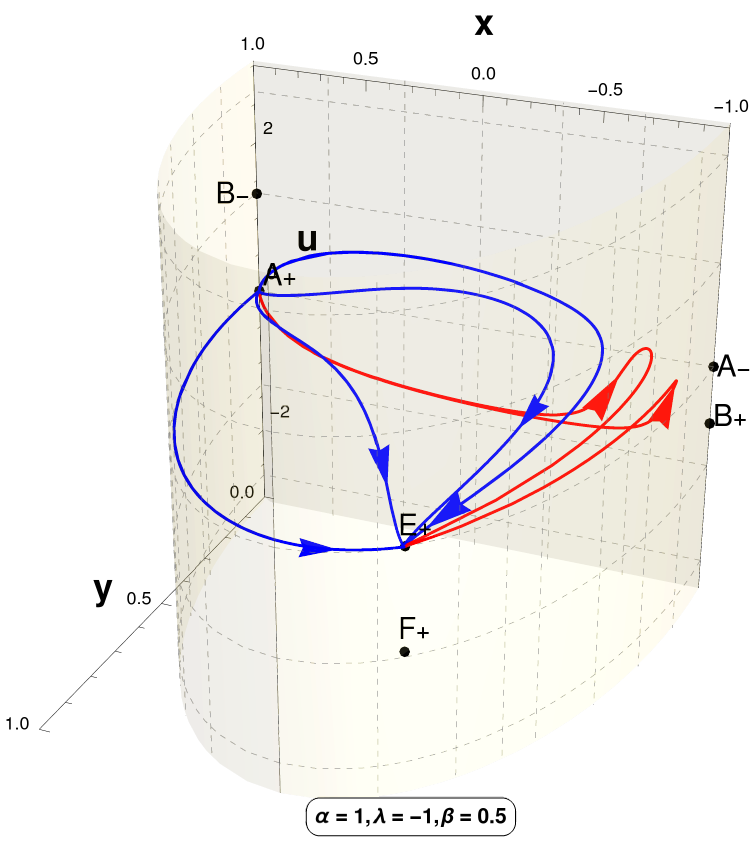}\quad \quad &
\includegraphics[width=0.465\textwidth]{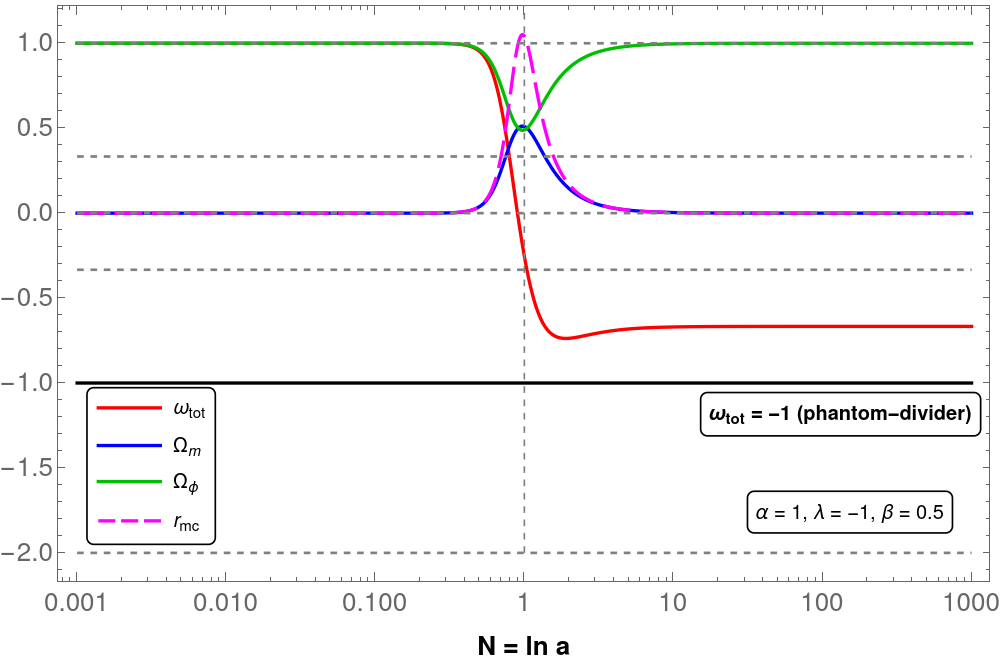}
\end{array}$}
\caption{Phase-space analysis and background evolution for \( \beta = 0.5 \). 
\textbf{Left:} 3D phase-space trajectories in \( (x, y, u) \) for \( \alpha = 1, \lambda = -1 \), showing  $E_{+}$ as a stable attractor. 
 \textbf{Right:} Evolution of key cosmological parameters: \( \omega_{\text{tot}} \), \( \Omega_m \), \( \Omega_\phi \), and \( r_{\text{mc}} \) as functions of \( N = \ln a \).}
\label{fig:5}
\end{figure}

Left panel of fig.~\ref{fig:5} illustrates the three-dimensional phase space evolution of the dynamical system in terms of the variables \( (x, y, u) \), corresponding respectively to the scalar field’s kinetic contribution, potential contribution, and the logarithmic growth rate of matter perturbations. The plot is drawn for the benchmark model parameters \( \alpha = 1, \lambda = -1, \beta = 0.5 \), where the dark energy--dark matter interaction is curvature-modulated and dynamically regulated to activate in the late universe. In this figure, the blue trajectories represent attractor solutions, while the red trajectories are repelled from unstable fixed points. The dynamical flows begin near early-time saddle or unstable nodes such as the kinetic or scaling solutions (\( A_\pm, B_\pm \)), where both the scalar field kinetic energy and matter energy densities are significant, and then evolve toward a late-time stable attractor (\( E_+ \)), where \( x \to \frac{\lambda}{\sqrt{6}} \), \( y \to -\sqrt{1-\frac{\lambda^2}{6}} \), and \( u \to 0 \). This attractor signifies a scalar-field dominated phase with negligible perturbation growth, consistent with cosmic acceleration. The phase flow behavior confirms that this benchmark supports a viable cosmological evolution: the interaction remains suppressed during the matter-dominated era, ensuring successful structure formation, while it becomes active at late times, allowing the universe to asymptotically approach an accelerated, dark energy-dominated phase. The clear separation between repelling and attracting trajectories in the phase space, guided by the curvature-screened interaction term, demonstrates the robustness of this model’s stability and its ability to interpolate naturally between early-time matter domination and late-time acceleration.\\

Right panel of fig.~\ref{fig:5} shows the cosmological evolution of key dynamical quantities as a function of the number of e-folds \( N = \ln a \), for the benchmark model with \( \alpha = 1 \), \( \lambda = -1 \), and \( \beta = 0.5 \), representing a moderately coupled and strongly curvature-modulated dark sector interaction. The plotted variables include the total equation of state parameter \( \omega_{\rm tot} \) (red), the energy densities of matter \( \Omega_m \) (blue) and the scalar field \( \Omega_\phi \) (green), and the matter-to-scalar field density ratio \( r_{\rm mc} = \Omega_m / \Omega_\phi \) (magenta dashed). In contrast to typical coupled dark energy models, the universe begins in a phase where \( \Omega_m \approx 1 \) and \( \Omega_\phi \approx 0 \), consistent with a standard matter-dominated epoch. This behavior is enforced by the exponential curvature modulation term, which suppresses the interaction when the Ricci scalar \( R \) is large (i.e., during early times when \( N \ll 1 \)), effectively delaying energy transfer from the scalar field to dark matter. As the universe expands and curvature drops, the interaction term becomes dynamically activated near \( N \sim 1 \), initiating a rapid transition in which \( \Omega_m \) and \( \Omega_\phi \) become comparable. This phase is captured by a peak in the ratio \( r_{\rm mc} \), signaling a temporary balancing of the two dark sectors. Following this transition, the scalar field takes over the energy budget with \( \Omega_\phi \to 1 \) and \( \Omega_m \to 0 \), while the equation of state drops toward a constant negative value \( \omega_{\rm tot} \to -0.7 \), indicating the onset of non-phantom type accelerated expansion.  The phantom divide at $\omega_{\rm tot} = -1$ is shown by the solid black line, and the evolution of the total EoS parameter stays entirely above it, confirming the absence of phantom crossing. The smooth yet sharp nature of this transition, along with the controlled late-time evolution, highlights the efficacy of the curvature-modulated interaction in naturally resolving the coincidence problem while ensuring consistency with both early-time structure formation and late-time cosmic acceleration.

\subsubsection*{\textbf{Case II:} \boldmath$\boldsymbol{\alpha = 2,\ \lambda = 1,\ \beta = 0.05}$}
 
\begin{figure}[H]
\centerline{$\begin{array}{cc}
\includegraphics[width=0.465\textwidth]{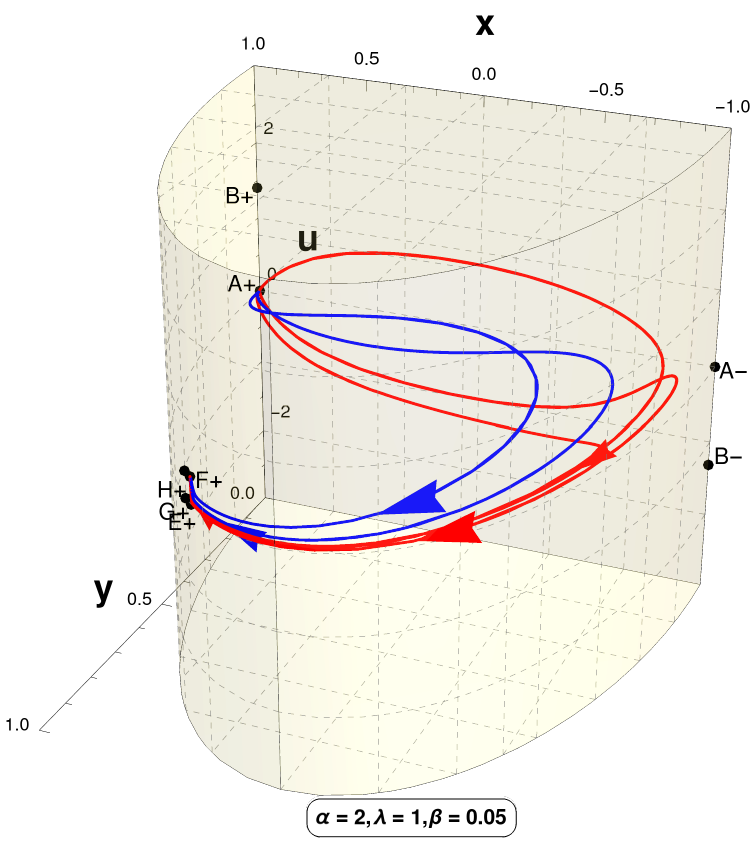}\quad \quad &
\includegraphics[width=0.465\textwidth]{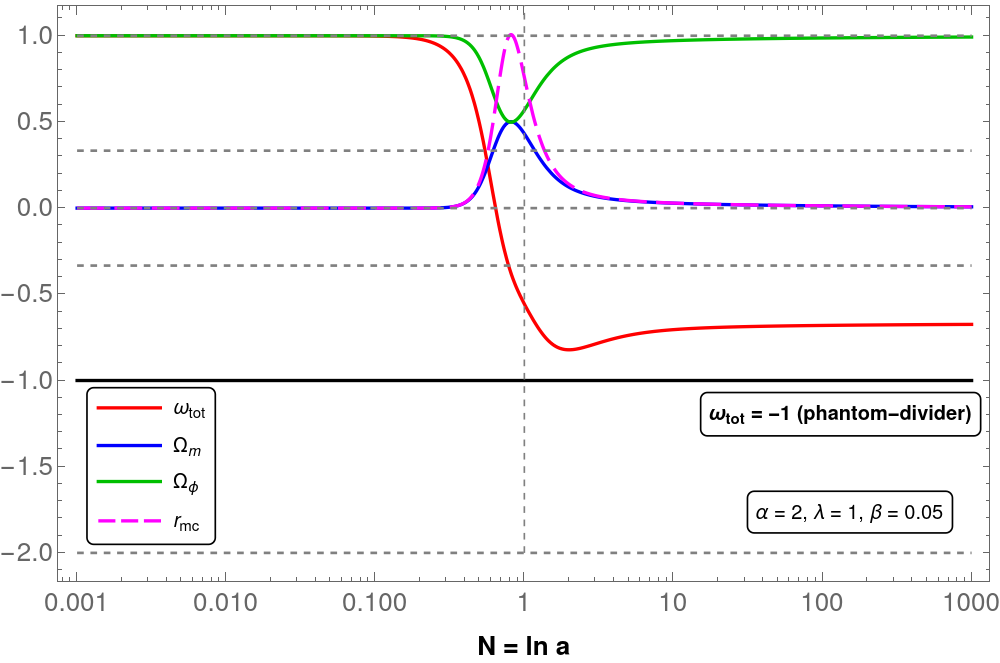}
\end{array}$}
\caption{Phase-space analysis and background evolution for \( \beta = 0.05 \). 
\textbf{Left:} 3D phase-space trajectories in \( (x, y, u) \) for \( \alpha = 2, \lambda = 1 \), showing  $F_{+}$ as a stable attractor. 
 \textbf{Right:} Evolution of key cosmological parameters: \( \omega_{\text{tot}} \), \( \Omega_m \), \( \Omega_\phi \), and \( r_{\text{mc}} \) as functions of \( N = \ln a \).}
\label{fig:6}
\end{figure}

Left panel of fig.~\ref{fig:6} presents the three-dimensional phase space evolution of the autonomous dynamical system in terms of the variables \( (x, y, u) \), where \( x \) denotes the normalized time derivative of the scalar field, \( y \) characterizes the contribution from the scalar field potential, and \( u \) represents the logarithmic growth rate of matter perturbations. This phase space corresponds to the benchmark scenario with model parameters \( \alpha = 2 \), \( \lambda = 1 \), and \( \beta = 0.05 \), describing a strongly coupled yet mildly curvature-modulated interaction between dark energy and dark matter. In the plot, the red trajectories are repelled from unstable nodes or saddle points such as \( A_\pm, B_\pm, E_+, G_+, H_+ \), while the blue trajectories converge toward the unique late-time attractor point labeled \( F_+ \). This critical point lies at finite values in the phase space and is characterized by the coordinates \( x = \lambda/\sqrt{6} \), \( y = \sqrt{1 - \lambda^2/6} \), and a nonzero matter growth rate \( u = \frac{1}{2} \left( -4 + 2 \alpha \lambda + \lambda^2 + \alpha \beta (-4 + \lambda^2) \right) \). This expression reflects the interplay between the coupling strength \( \alpha \), the slope of the scalar field potential \( \lambda \), and the curvature modulation \( \beta \), all of which regulate the decay or persistence of matter perturbations. The dynamical evolution begins near matter- or kinetic-dominated regions, where structure growth is active (reflected in positive \( u \)), and proceeds toward the attractor \( F_+ \), where \( u \) asymptotes to a constant, typically small or negative, indicating a cessation or suppression of structure growth. This behavior confirms that the benchmark supports a viable transition from a growth-permitting early universe to a late-time scalar-field dominated accelerating phase, with dynamically controlled interaction. The presence of only one globally stable critical point surrounded by unstable nodes highlights the predictiveness and robustness of this scenario in guiding cosmological evolution toward a well-defined attractor, effectively linking structure formation and dark energy domination within a unified dynamical framework.

Right panel of fig.~\ref{fig:6} illustrates the evolution of key cosmological quantities as functions of the e-folding number \( N = \ln a \), for the benchmark model with \( \alpha = 2 \), \( \lambda = 1 \), and \( \beta = 0.05 \), representing a strongly coupled but weakly curvature-modulated dark energy--dark matter interaction. The curves correspond to the total effective equation of state \( \omega_{\rm tot} \) (red), the energy densities of matter \( \Omega_m \) (blue) and scalar field \( \Omega_\phi \) (green), and the matter-to-dark energy density ratio \( r_{\rm mc} = \Omega_m / \Omega_\phi \) (magenta dashed). At very early times (\( N \lesssim 0.1 \)), the scalar field dominates the cosmic energy budget with \( \Omega_\phi \gg \Omega_m \), even though the interaction term is strongly suppressed by large curvature through the Ricci-modulated exponential coupling. In this regime, the scalar field evolves nearly independently, and its dynamics effectively mimic a dark energy-like component, driving expansion in a way that deviates from standard matter domination. As the universe expands, the matter energy density grows relative to the scalar field, and around \( N \sim 1 \), the system transitions into a quasi-scaling regime where \( \Omega_m \sim \Omega_\phi \). This is reflected in the pronounced peak in the ratio \( r_{\rm mc} \), which indicates a period of near balance between the dark sectors-a behavior with important implications for addressing the coincidence problem. At late times (\( N \gg 1 \)), the system settles into a dark energy-dominated attractor, with \( \Omega_\phi \to 1 \), \( \Omega_m \to 0 \), and \( \omega_{\rm tot} \to -0.65 \), ensuring persistent cosmic acceleration.   The solid black line denotes the phantom divide at $\omega_{\rm tot} = -1$, while here evolution of total EoS parameter stays above this threshold, confirming non-phantom type acceleration. Overall, the evolution reflects the key phenomenology of curvature-modulated interaction models: a suppressed coupling in the high-curvature early universe that protects structure formation, an intermediate balancing phase between dark matter and dark energy, and a stable transition to late-time non-phantom type acceleration driven by the scalar field.

\subsection{Variation of coincidence and total EoS parameter}
\begin{figure}[H]
\centering
        \includegraphics[width=0.75\textwidth]{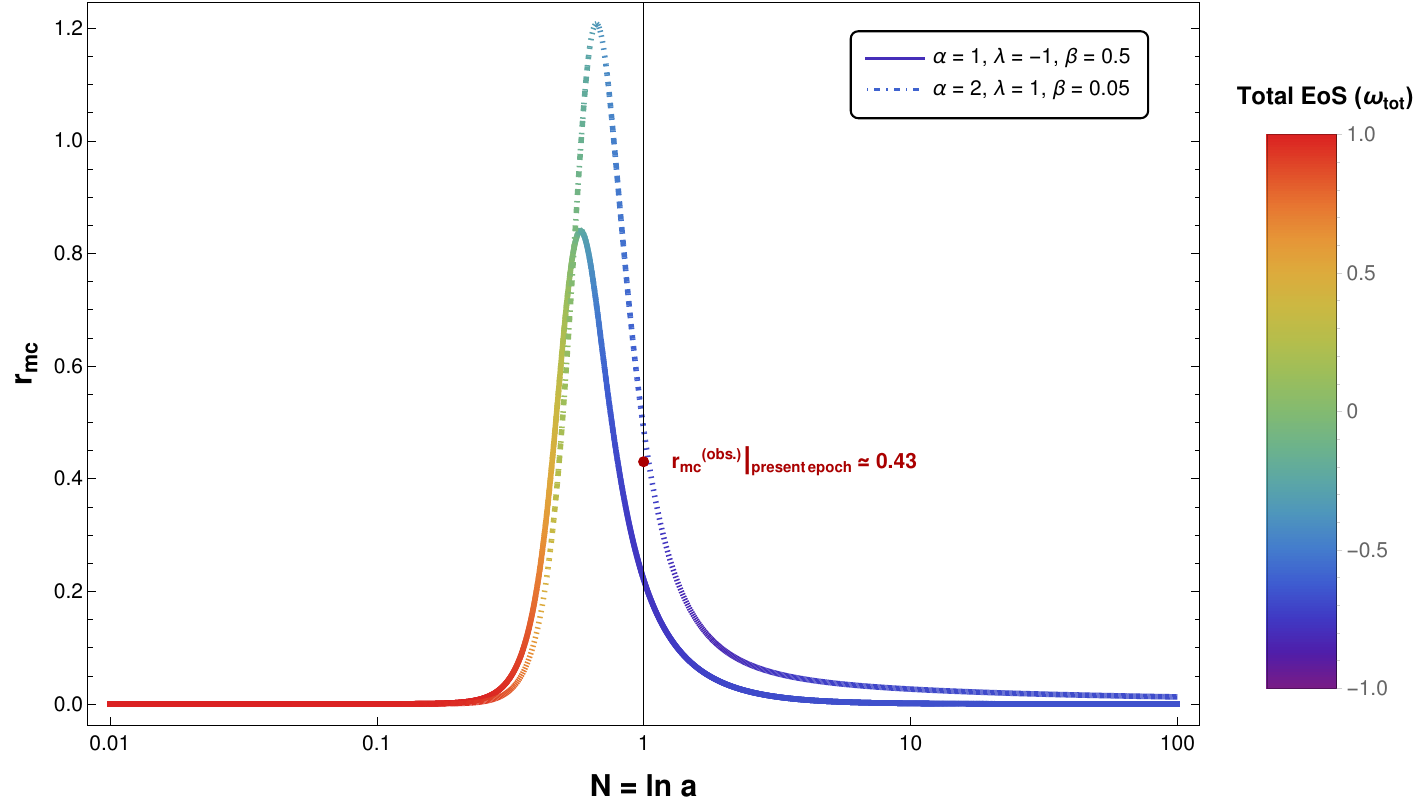}
        \caption{Evolution of the coincidence parameter \( r_{\text{mc}} \) vs. \( N = \ln a \) for 
\( (\alpha=1, \lambda=-1, \beta=0.5) \) (solid) and \( (\alpha=2, \lambda=1, \beta=0.05) \) (dashed). 
Color bar indicates the variation of total EoS parameter  (\( \omega_{\text{tot}} \)).}
        \label{fig:7}
\end{figure}

In order to understand the late-time behavior and assess the severity of the cosmic coincidence problem within our interacting dark energy–dark matter model, we analyze the evolution of the coincidence parameter \( r_{\rm mc} = \frac{\Omega_m}{\Omega_\phi} \), alongside the total effective equation of state \( \omega_{\text{tot}} \), for two representative benchmark cases: (i) \( \alpha = 1, \lambda = -1, \beta = 0.5 \) and (ii) \( \alpha = 2, \lambda = 1, \beta = 0.05 \). The interaction term \( Q_0 = \alpha \kappa \rho_m \dot{\phi} \left(1 - \frac{\beta}{6H^2} R\right) \) induces energy exchange between the dark sectors, with the curvature-modulating parameter \( \beta \) controlling the strength of this coupling. From the plotted evolution in Fig.~\ref{fig:7}, both models reproduce the standard cosmological sequence, with early-time stiff matter behavior (\( \omega_{\text{tot}} \sim 1 \)), followed by a radiation-like phase (\( \omega_{\text{tot}} \approx \frac{1}{3} \)), then matter domination (\( \omega_{\text{tot}} \approx 0 \)), and eventually a transition to accelerated expansion with \( \omega_{\text{tot}} < -\frac{1}{3} \). Benchmark (i) shows a smoother and earlier transition into the accelerating regime, whereas benchmark (ii) exhibits a sharper transition driven by the steeper potential slope (\( \lambda = 1 \)). The evolution of \( r_{\rm mc} \) reveals further distinctions: in benchmark (i), \( r_{\rm mc} \) gradually decreases and stabilizes near a small constant value at late times, indicating prolonged coexistence of dark matter and dark energy and thereby offering a dynamical alleviation of the coincidence problem. In contrast, benchmark (ii) leads to a rapid drop in \( r_{\rm mc} \), suggesting dark energy domination without a significant intermediate tracking regime. Remarkably, the present-day observed value of the coincidence parameter, \( r_{\rm mc}^{\text{(obs.)}} \approx 0.43 \), is also well captured by both benchmark evolutions, providing an important consistency check for the model. The ability of the model to maintain \( r_{\rm mc} \sim \mathcal{O}(1) \) over an extended epoch is a crucial feature for addressing the coincidence problem, which questions why dark matter and dark energy densities are of comparable magnitude today. The curvature-sensitive interaction governed by \( \beta \), together with the choice of \( \alpha \) and \( \lambda \), thus plays a pivotal role in achieving a viable cosmological trajectory that not only leads to late-time acceleration but also dynamically reduces the degree of fine-tuning typically required in $\Lambda$CDM models.

\subsection{Evolution of growth rate, density contrast and index parameter}
\begin{figure}[H]
\centerline{$\begin{array}{cc}
\includegraphics[width=0.47\textwidth,height=0.35\textwidth]{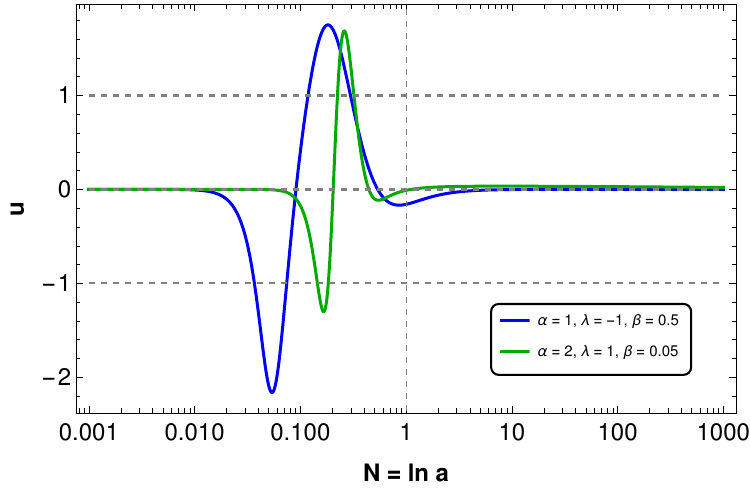} &
\includegraphics[width=0.47\textwidth,height=0.35\textwidth]{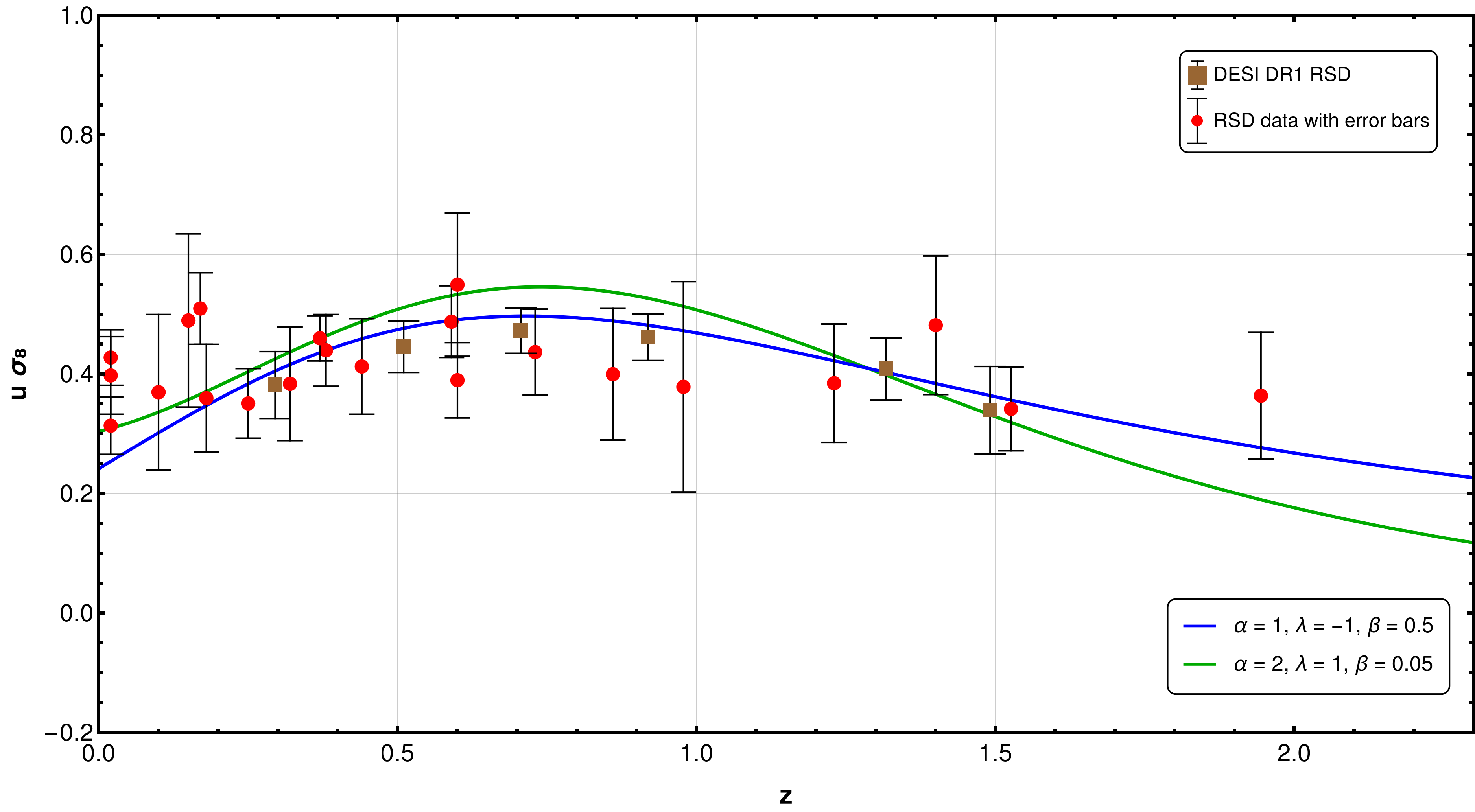}
\end{array}$}
\caption{Evolution of growth rate observables in two benchmark points of interacting DE-DM models. 
\textbf{Left:} Growth rate of matter perturbations \( u \) as a function of e-folding number \( N = \ln a \). 
 \textbf{Right:} Redshift evolution of the observable \( u\sigma_8(z) \), compared with  DESI DR1 RSD data  including error bars (brown points) \cite{DESI:2024jxi} and RSD data including error bars (red points)\cite{Sagredo:2018ahx,Kazantzidis:2018rnb}. Curves correspond to \( (\alpha=1, \lambda=-1, \beta=0.5) \) (blue) and \( (\alpha=2, \lambda=1, \beta=0.05) \) (green).}
\label{fig:8}
\end{figure}

 Left panel of fig.~\ref{fig:8} shows the evolution of the growth rate variable \( u \), defined as a dimensionless measure of the growth of matter perturbations, plotted as a function of the e-folding number \( N = \ln a \) for two benchmark parameter sets: (i) \( \alpha = 1, \lambda = -1, \beta = 0.5 \) and (ii) \( \alpha = 2, \lambda = 1, \beta = 0.05 \). The variable \( u \), introduced as part of the dynamical system analysis (DSA), effectively captures the normalized rate of change of the matter density contrast and provides direct insight into the efficiency of structure formation within the interacting dark sector framework. For both benchmarks, the evolution of \( u \) shows a transition from a positive to negative regime as the universe evolves from early to late times. At early times (low \( N \)), both models exhibit \( u > 0 \), indicating growing modes consistent with the standard matter-dominated epoch, where gravitational instability leads to the formation of large-scale structures. This growth is particularly pronounced for benchmark (ii), with \( \alpha = 2 \), suggesting that a stronger coupling enhances the initial perturbation growth due to more efficient energy transfer from the scalar field to dark matter.  As the universe enters the dark energy-dominated regime (large \( N \)), \( u \) gradually decreases and eventually becomes negative, reflecting a suppression of perturbation growth. This behavior is expected, as the accelerating expansion driven by the scalar field counteracts gravitational collapse. The curvature-modulated interaction plays a crucial role in this transition: the term \( \left(\frac{\beta R}{6H^2} \right) \) dynamically regulates the interaction strength such that it is suppressed at early times (when curvature is high) and becomes effective at late times. In benchmark (i), with higher \( \beta = 0.5 \), this modulation becomes more prominent, leading to a sharper decline in \( u \) compared to benchmark (ii), where \( \beta = 0.05 \). Thus, curvature feedback not only impacts the background expansion but also directly modifies the growth history of matter fluctuations. The late-time negative values of \( u \) signal the onset of the accelerated expansion phase, where the suppression of structure formation aligns with observational evidence from weak lensing and galaxy clustering data. This transition is also important for addressing the cosmic coincidence problem, as the dynamically screened interaction allows for a prolonged matter-dominated era followed by a smooth entry into acceleration, with the perturbation growth naturally tapering off. Hence, the behavior of \( u(N) \) in these models provides a consistent picture that links background dynamics, interaction physics, and structure formation, highlighting the capability of the curvature-modulated coupling to yield observationally viable cosmologies with testable signatures.\\

 Right panel of Fig.~\ref{fig:8} illustrates the redshift evolution of the growth observable $u\sigma_8(z)$ for two benchmark parameter sets within the curvature-modulated interacting quintessence model: (i) $\alpha = 1, \lambda = -1, \beta = 0.5$ (blue curve), and (ii) $\alpha = 2, \lambda = 1, \beta = 0.05$ (green curve). These predictions are compared with recent DESI DR1 RSD growth constraints \cite{DESI:2024jxi} (brown squares with error bars)  together with the compilation of redshift-space distortion (RSD) measurements (red points with error bars) \cite{Sagredo:2018ahx,Kazantzidis:2018rnb}. The observable $u\sigma_8$ encodes both the growth rate of matter perturbations and the amplitude of clustering on $8\,h^{-1}\,\mathrm{Mpc}$ scales, thereby serving as a direct probe of cosmic structure formation. Both interacting models reproduce the general observed behavior: an increase of $u\sigma_8$ at intermediate redshifts, peaking around $z \sim 0.8$, followed by a gradual decline at higher redshifts. Benchmark (ii), characterized by stronger coupling ($\alpha = 2$) and weaker curvature modulation ($\beta = 0.05$), exhibits a more pronounced peak and an earlier turnover, pointing to enhanced growth at earlier epochs due to more efficient energy transfer from dark energy to dark matter. In contrast, benchmark (i), with milder coupling and stronger curvature modulation ($\beta = 0.5$), yields a flatter evolution with suppressed growth at intermediate redshifts, consistent with delayed activation of the interaction. Both benchmarks remain consistent with the full set of observational constraints, with benchmark (ii) showing slightly closer alignment with the DESI DR1  points in the range $0.5 < z < 1.0$. These findings highlight the role of curvature-modulated DE–DM interactions in shaping the growth history of matter perturbations, providing testable departures from $\Lambda$CDM while remaining compatible with current RSD and DESI measurements.\\

\begin{figure}[H]
\centerline{$\begin{array}{cc}
\includegraphics[width=0.47\textwidth,height=0.34\textwidth]{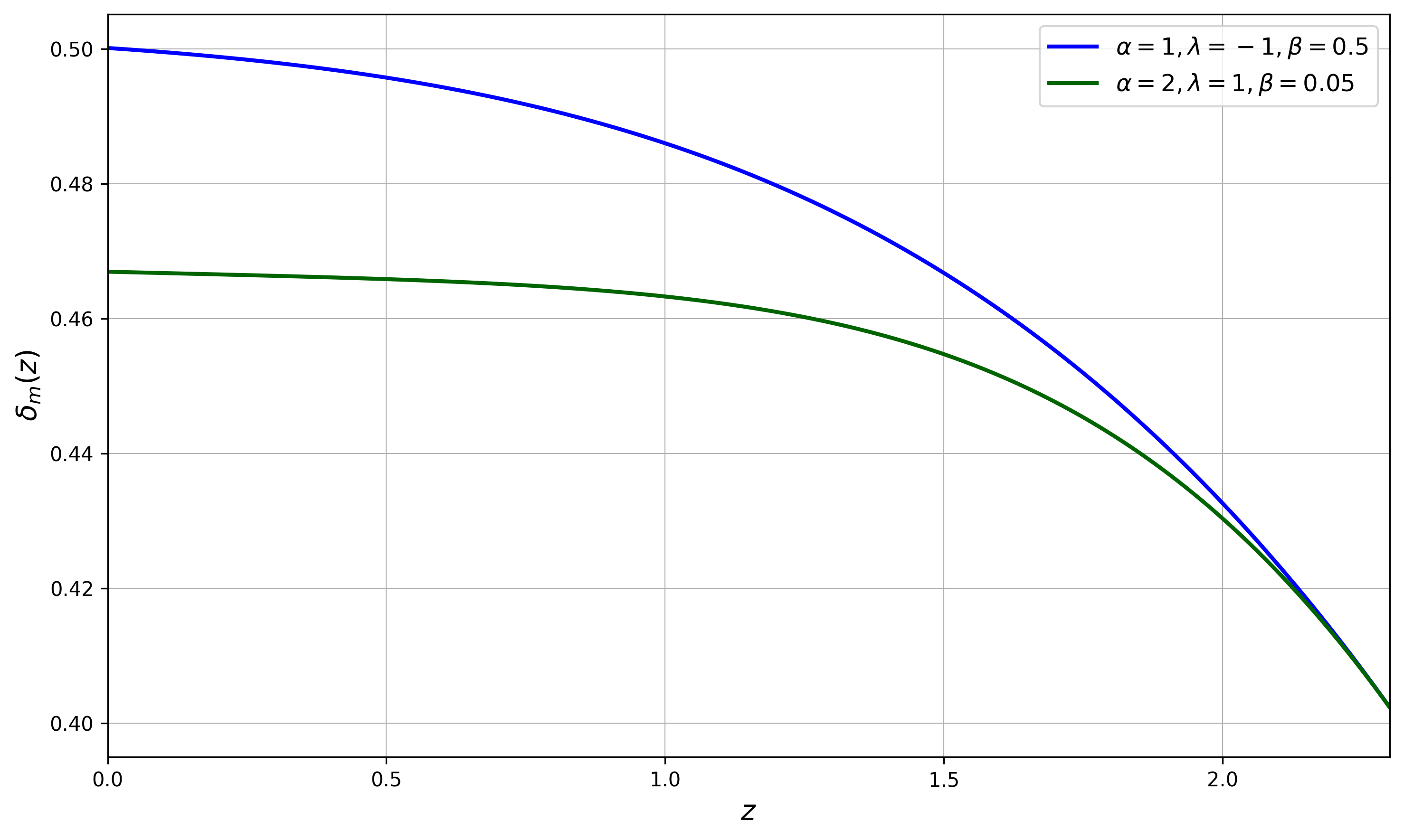}  &
\includegraphics[width=0.47\textwidth,height=0.34\textwidth]{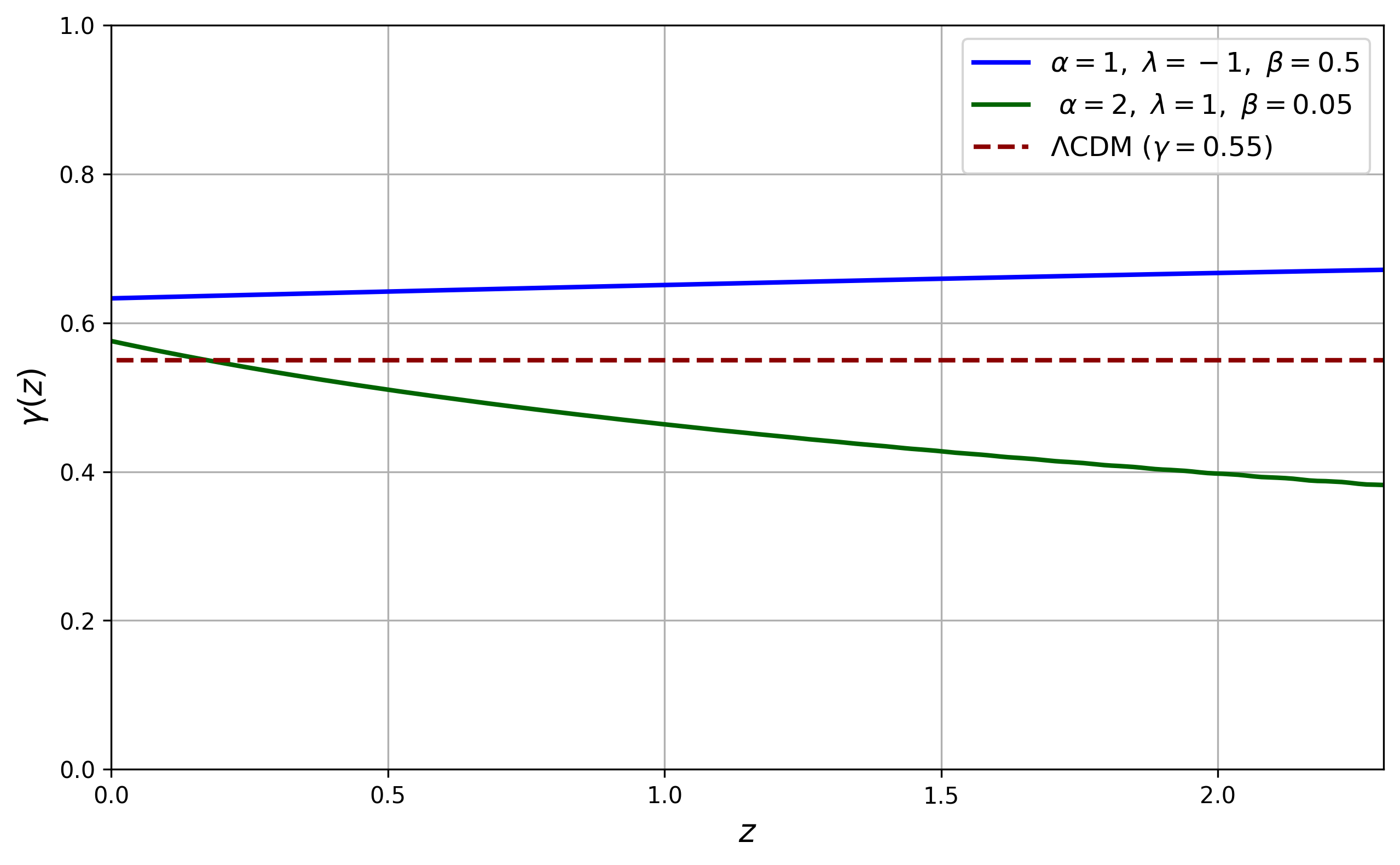}
\end{array}$}
\caption{Evolution of growth-related parameters for two interacting DE-DM models. 
\textbf{Left:} Matter density contrast \( \delta_m(z) \). 
\textbf{Right:} Growth index \( \gamma(z) \). 
Shown for \( (\alpha=1, \lambda=-1, \beta=0.5) \) (blue), 
\( (\alpha=2, \lambda=1, \beta=0.05) \) (green), and \(\Lambda\)CDM (dashed red, \( \gamma = 0.55 \)).}
\label{fig:9}
\end{figure}

In the left panel of fig.~\ref{fig:9} presents the redshift evolution of the matter density contrast \( \delta_m(z) \), obtained by numerically solving the autonomous system of eqs.(\ref{eq:dx})–(\ref{eq:dlm}) governing the variables \( x, y,u, \delta_m \) with appropriate initial conditions. The two curves correspond to benchmark parameter sets: (i) \( \alpha = 1, \lambda = -1, \beta = 0.5 \) (blue) and (ii) \( \alpha = 2, \lambda = 1, \beta = 0.05 \) (green). The variable \( \delta_m \) quantifies the amplitude of matter perturbations and serves as a key indicator of structure formation in the universe. As evident from the figure, both models exhibit a gradual suppression of \( \delta_m \) with increasing redshift, consistent with the expectation that structure growth slows in the past as the universe becomes increasingly dominated by dark energy. Benchmark (i), which features a stronger curvature modulation (\( \beta = 0.5 \)) and a less steep potential (\( \lambda = -1 \)), shows higher values of \( \delta_m \) across the redshift range, indicating more sustained structure growth. This can be attributed to the enhanced late-time activation of the interaction term \( Q_0 = \alpha \kappa \rho_m \dot{\phi} \left(1 - \frac{\beta R}{6H^2} \right) \), which becomes dynamically significant when the curvature \( R \) is low and thereby facilitates delayed but efficient matter clustering. In contrast, benchmark (ii), with a weaker modulation and a steeper potential, predicts suppressed growth due to earlier dark energy domination and reduced coupling strength at late times. Cosmologically, the behavior of \( \delta_m(z) \) reflects how energy exchange between dark matter and dark energy influences the amplitude and timing of structure formation. Higher late-time \( \delta_m \) values, as seen in benchmark (i), suggest better consistency with observed RSD  and galaxy clustering data. From a model prediction standpoint, this implies that curvature-modulated interactions can regulate the growth history in a way that not only aligns with current observations but also offers a dynamical mechanism to address late-time suppression in structure formation. Thus, the figure reinforces that the growth of matter density contrast is highly sensitive to both the interaction strength \( \alpha \) and the modulation parameter \( \beta \), and underscores the predictive capability of the model in describing structure formation within a dynamically evolving dark sector.

In the right panel of fig.~\ref{fig:9} illustrates the evolution of the growth index \( \gamma(z) \), defined as
\begin{equation}
\gamma(z) \approx \frac{\ln u(z)}{\ln \Omega_m(z)},
\end{equation}
where \( u(z) \equiv \frac{d\ln \delta_m}{d\ln a} \) is the linear growth rate of matter perturbations and \( \Omega_m(z) \) is the critical-matter density parameter. The quantity \( \gamma \) encapsulates the sensitivity of the growth rate to the background expansion and is widely used as a diagnostic for distinguishing modified gravity and dark energy models from the standard \(\Lambda\)CDM cosmology, which yields a nearly constant value \( \gamma \approx 0.55 \) in the matter-dominated era (shown here as the dashed red line). The plot compares \( \gamma(z) \) for two interacting curvature-modulated quintessence models: benchmark (i) with \( \alpha = 1, \lambda = -1, \beta = 0.5 \) (blue curve), and benchmark (ii) with \( \alpha = 2, \lambda = 1, \beta = 0.05 \) (green curve). The interaction modifies the continuity and Euler equations for dark matter, thereby altering the growth rate \( u(z) \) and consequently the growth index. In the interacting scenario with stronger curvature modulation in benchmark (i), \( \gamma(z) \) remains higher than \(\Lambda\)CDM throughout, slowly increasing from \( \sim 0.62 \) at \( z = 0 \) to \( \sim 0.65 \) at higher redshift. This indicates a slower growth of structures due to a late-time energy transfer from dark matter to dark energy, which suppresses clustering. In contrast, benchmark (ii) features a stronger coupling \( \alpha = 2 \) but weaker curvature modulation, leading to a sharper decline in \( \gamma(z) \) with redshift—from \( \sim 0.57 \) at \( z = 0 \) to below \( 0.4 \) by \( z \sim 2 \). This implies a more rapid growth of structures in the early universe, driven by enhanced energy flow from the scalar field to dark matter. The deviation from the \(\Lambda\)CDM prediction in both cases underscores the dynamical impact of interaction terms on structure formation. Mathematically, for the interacting model with \( Q_0 = \alpha \kappa \rho_m \dot{\phi} \left(1 - \frac{\beta R}{6H^2} \right) \), the modified perturbation evolution alters the friction and source terms in the perturbation equation for \( \delta_m \), thus changing \( u(z) \) and consequently \( \gamma(z) \). Physically, the growth index is particularly sensitive to the redshift-dependent strength of the interaction: at higher redshifts (large \( R \)), the interaction is suppressed due to the exponential screening, aligning with standard growth; at lower redshifts, as \( R \) decreases, the interaction activates, leading to suppressed or enhanced growth depending on the direction of energy flow. Therefore, the variation in \( \gamma(z) \) not only quantifies the deviation from \(\Lambda\)CDM but also encodes the physical imprint of the DE-DM interaction history, offering a powerful tool for testing coupled quintessence models against future redshift-dependent structure growth observations.

\subsection{Comparison with observational constraints}

\begin{figure}[H]
\centerline{$\begin{array}{cc}
\includegraphics[width=0.47\textwidth,height=0.34\textwidth]{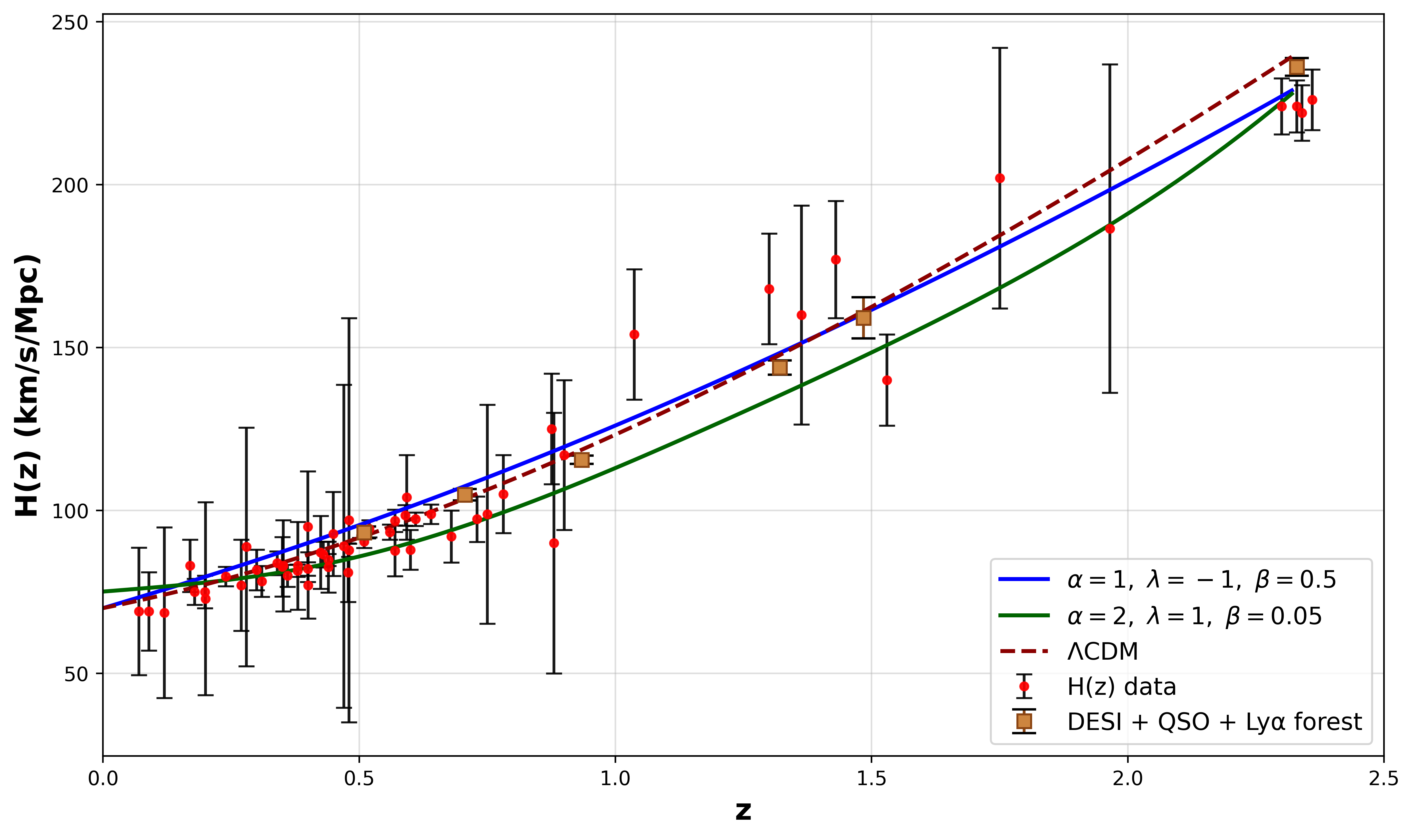}  &
\includegraphics[width=0.47\textwidth,height=0.34\textwidth]{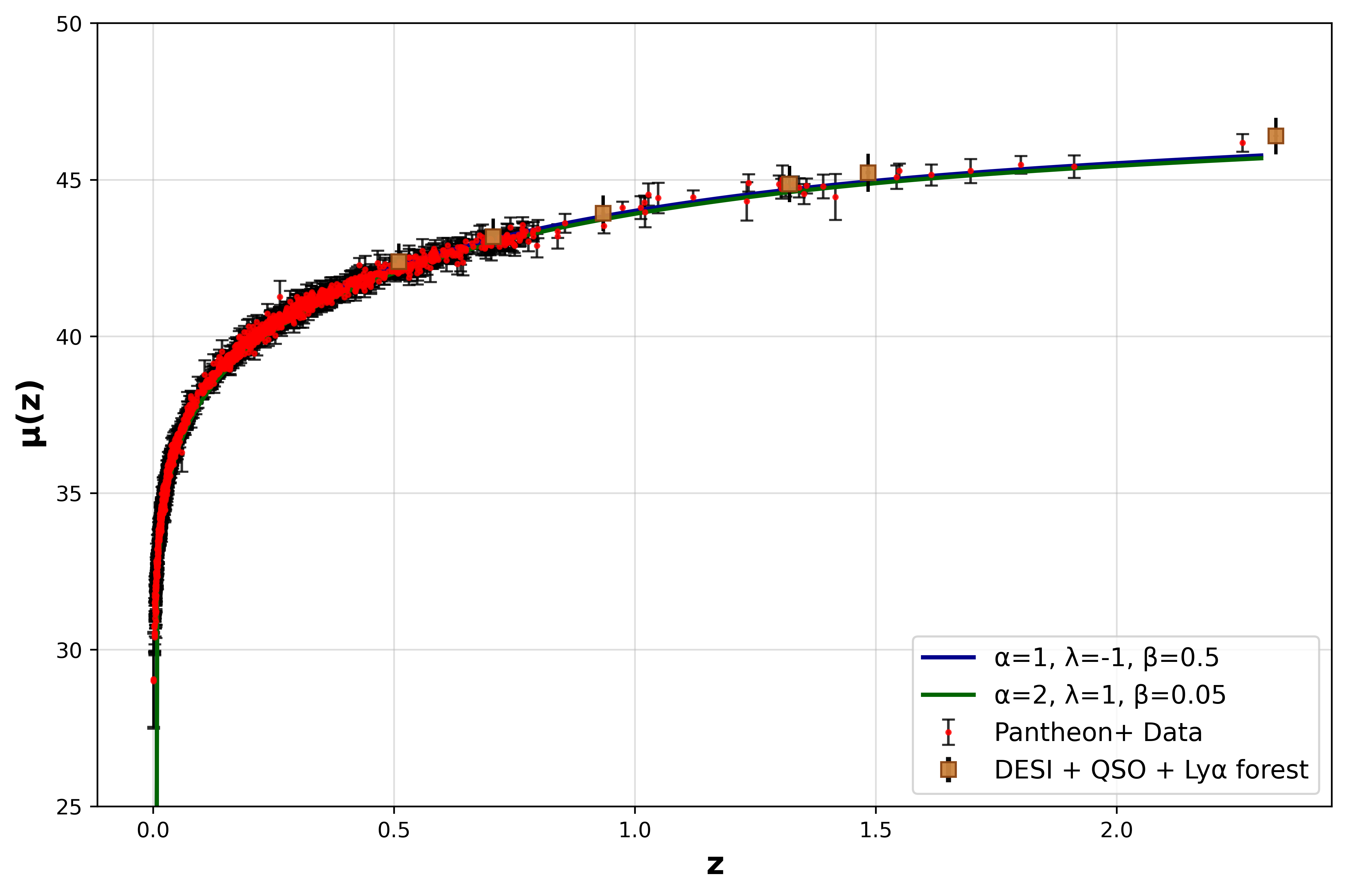}
\end{array}$}
\caption{Evolution of background observables for two types of interacting DE-DM models. 
\textbf{Left:} Hubble parameter \( H(z) \) compared with OHD  data-sets \cite{Gadbail:2024rpp} and DESI + QSO + Ly$\alpha$ forest data \cite{DESI:2025zgx}. 
\textbf{Right:} Distance modulus \( \mu(z) \) compared with Pantheon+ data \cite{Scolnic:2021amr} and DESI + QSO + Ly$\alpha$ forest data  \cite{DESI:2025zgx}. Curves shown for \( (\alpha=1, \lambda=-1, \beta=0.5) \) (blue), 
\( (\alpha=2, \lambda=1, \beta=0.05) \) (green), and \(\Lambda\)CDM (dashed red in left panel).}
\label{fig:10}
\end{figure}

 In the right panel of Fig.~\ref{fig:10}, we show the redshift evolution of the Hubble parameter $H(z)$, which quantifies the expansion rate of the universe, for two benchmark models within the curvature-modulated interacting quintessence scenario: (i) $\alpha = 1, \lambda = -1, \beta = 0.5$ (blue curve), and (ii) $\alpha = 2, \lambda = 1, \beta = 0.05$ (green curve). For reference, the standard $\Lambda$CDM prediction is plotted as a dashed red line, and observational Hubble data points \cite{Gadbail:2024rpp} (red dots with error bars) are also included. The Hubble parameter is derived from the Friedmann equation, with the interaction term  $Q_0 = \alpha \kappa \rho_m \dot{\phi} \left(1 - \tfrac{\beta R}{6H^2} \right)$ altering the background dynamics by modifying the evolution of the matter and scalar field densities. We additionally overlay the and DESI  + QSO + Ly$\alpha$ forest radial mode measurements \cite{DESI:2025zgx} ($d_H/r_d$), converted into $H(z)$ using $H(z) = \frac{c}{(d_H/r_d)\,r_d}$, shown as filled brown squares with black error bars at effective redshifts $z = 0.510, 0.706, 0.934, 1.321, 1.484$ and $2.330$.  These high-precision datasets provide robust constraints on the late-time expansion history and significantly strengthen the comparison with theoretical models. Both interacting benchmarks exhibit good consistency with the combined observational data across the redshift range $0 < z < 2.3$. Benchmark (i), with a larger curvature modulation ($\beta = 0.5$), tracks the $\Lambda$CDM curve closely at low to intermediate redshifts but predicts a slightly lower expansion rate at higher $z$, reflecting the delayed onset of the interaction. Benchmark (ii), featuring a stronger coupling ($\alpha = 2$) and smaller modulation parameter ($\beta = 0.05$), instead yields a mildly slower expansion at low redshifts before converging to the same high-$z$ trend. This behavior reflects the earlier activation of the interaction when the coupling strength dominates over curvature suppression. Overall, the comparison shows that both benchmark models deliver expansion histories consistent with the observational error bounds, including the new DESI measurements \cite{DESI:2025zgx}. The small but noticeable departures from $\Lambda$CDM-particularly at intermediate redshifts-indicate that curvature-modulated interactions imprint subtle, testable signatures on late-time expansion. Such deviations may play a role in alleviating the Hubble tension, as interacting models allow for a more flexible framework connecting early and late-universe determinations of $H_0$. Therefore, the $H(z)$ analysis confirms the observational viability of the proposed interacting quintessence models while highlighting the distinct dynamical effects introduced by curvature-regulated DE-DM coupling.

 In the right panel of Fig.~\ref{fig:10}, we present the redshift evolution of the distance modulus $\mu(z)$, defined as
\begin{equation}
\mu(z) = 5 \log_{10} \left[ \frac{d_L(z)}{ \mathrm{Mpc}} \right] +25
= 5 \log_{10} \left[ \frac{(1+z) c}{\mathrm{Mpc}} \int_0^z \frac{dz'}{H(z')} \right] + 25.
\end{equation}
where $d_L(z)$ is the luminosity distance and $H(z)$ is the Hubble parameter predicted by the models. This observable is compared with the Pantheon+ compilation of Type Ia supernovae \cite{Scolnic:2021amr} (red points with error bars), which covers the range $0 < z < 2.3$. In addition, we incorporate the DESI + QSO + Ly$\alpha$ forest data-sets transverse measurements \cite{DESI:2025zgx} of $d_M/r_d$. After multiplying by $r_d$ to obtain $d_M(z)$, we use the standard relation $d_L(z) = (1+z) d_M(z)$ to convert to luminosity distances, and subsequently derive the distance moduli. The resulting points are shown as brown squares with black error bars at effective redshifts $z = 0.510, 0.706, 0.934, 1.321, 1.484$ and $2.330$. These high-precision data points provide independent, high-precision constraints on the background expansion and complement the supernova dataset. The figure includes theoretical predictions for two benchmark parameter sets of the curvature-modulated interacting quintessence scenario: (i) $\alpha = 1, \lambda = -1, \beta = 0.5$ (blue curve), and (ii) $\alpha = 2, \lambda = 1, \beta = 0.05$ (green curve). Both models show excellent agreement with the Pantheon+ data and the DESI  constraints across the entire redshift interval, thereby confirming the compatibility of curvature-modulated coupled quintessence with current luminosity distance measurements. The small deviations between the two curves illustrate how late-time expansion is sensitive to the interaction strength $\alpha$, the scalar potential slope $\lambda$, and the curvature modulation parameter $\beta$. Benchmark (i), with stronger curvature modulation ($\beta = 0.5$), leads to a sharper late-time activation of the interaction, slightly shifting the expansion at higher redshifts. Benchmark (ii), with a larger coupling ($\alpha = 2$) but weaker modulation ($\beta = 0.05$), produces an expansion history that is nearly degenerate with benchmark (i), pointing to parameter degeneracy at the background level. The successful consistency with both supernova and DESI + QSO + Ly$\alpha$ forest  datasets demonstrates that the proposed curvature-modulated interacting quintessence framework remains observationally viable for background cosmology, while leaving room for distinguishing signatures at the perturbation level.

\subsection{Kinematic evolution: deceleration and jerk Parameter}

\begin{figure}[H]
\centerline{$\begin{array}{cc}
\includegraphics[width=0.47\textwidth,height=0.35\textwidth]{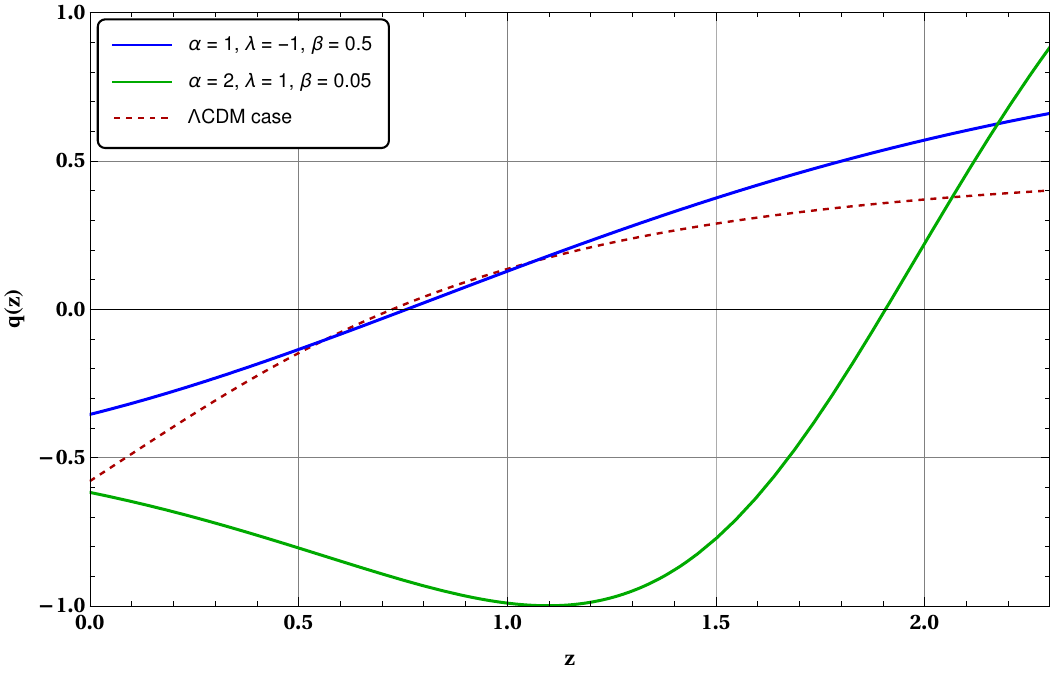}  &
\includegraphics[width=0.47\textwidth,height=0.35\textwidth]{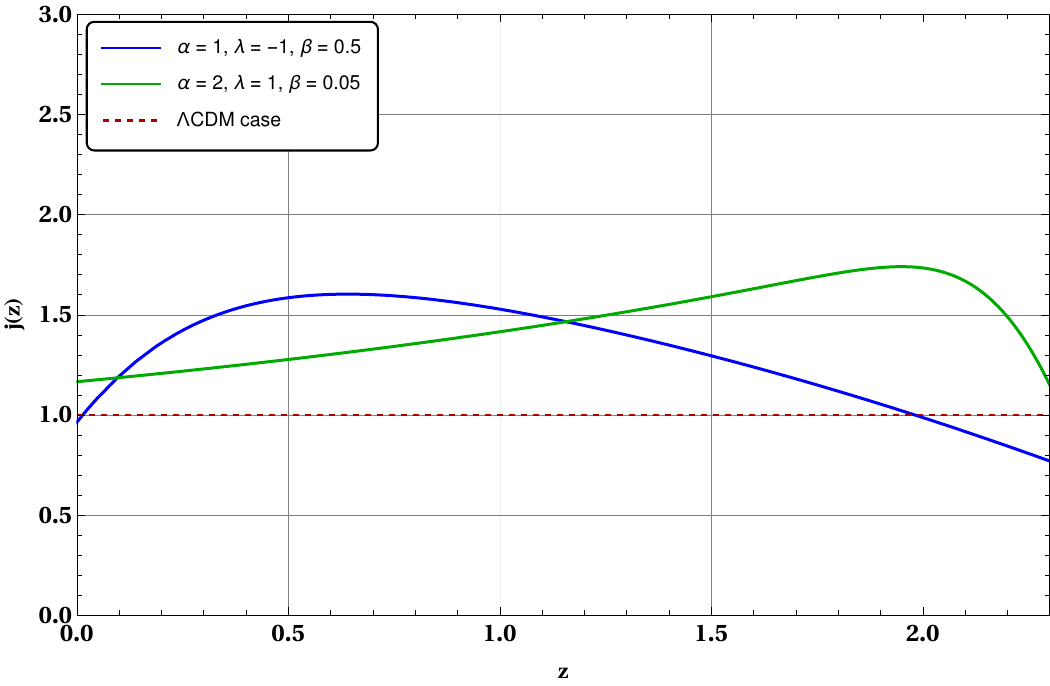}
\end{array}$}
\caption{Evolution of cosmographic parameters for two interacting DE--DM models. 
\textbf{Left:} Deceleration parameter \( q(z) \). 
\textbf{Right:} Jerk parameter \( j(z) \). 
Shown for \( (\alpha=1, \lambda=-1, \beta=0.5) \) (blue), 
\( (\alpha=2, \lambda=1, \beta=0.05) \) (green), and \(\Lambda\)CDM (dashed red).}
\label{fig:11}
\end{figure}

In the both panels of  fig.~\ref{fig:11} provide an in-depth look at the cosmographic evolution of the universe in the context of the curvature-modulated interacting quintessence model, using the deceleration parameter \( q(z) \) and the jerk parameter \( j(z) \) as key diagnostics. These kinematical quantities, previously defined, are derived from the first and second derivatives of the Hubble parameter and offer a powerful means of distinguishing cosmological models without relying on specific assumptions about matter content.  Evolution of the deceleration parameter in fig.~\ref{fig:11} reveals that both interacting benchmark models undergo a smooth transition from a decelerating to an accelerating expansion phase, which is consistent with the transition epoch inferred from multiple observational probes. Between the two cases, benchmark (ii), with stronger coupling (\( \alpha = 2 \)) and weaker curvature modulation (\( \beta = 0.05 \)), transitions to acceleration (at $z \sim 1.9$) slightly earlier than benchmark (i), which has a weaker coupling (\( \alpha = 1 \)) and stronger modulation (\( \beta = 0.5 \)) (at $z \sim 0.7$). This earlier transition is expected, as a stronger interaction enhances the energy transfer from dark energy to dark matter, promoting earlier dominance of the dark energy component. Deceleration characteristic of benchmark (i)
closely resembles that of the $\Lambda$CDM scenario.  In the right panel of fig.~\ref{fig:11} further illustrates the behavior of the jerk parameter, which remains constant at unity for the \(\Lambda\)CDM model. In contrast, both interacting models exhibit mild deviations from this constant value, particularly over the intermediate redshift range. The deviations are more significant for benchmark (ii), reflecting the more dynamic nature of the interaction at earlier times due to the relatively low curvature suppression. These departures from \( j = 1 \) are a direct consequence of the evolving interaction term, especially the curvature-modulated factor that alters the acceleration rate of the universe in a redshift-dependent way. Together, the behavior of \( q(z) \) and \( j(z) \) in both benchmarks confirms the observational consistency of the interacting model while highlighting its predictive differences from \(\Lambda\)CDM. The deviations in the jerk parameter, though modest, serve as a distinctive signature of curvature-modulated interactions and can be tested with next-generation high-precision cosmographic data. These cosmographic features complement the background and perturbation analyses, reinforcing the viability of both benchmark choices while emphasizing their distinct dynamical imprints.

\section{Conclusion}
\label{con}

In this work, we have investigated a class of interacting dark energy–dark matter models in which the interaction is introduced via a curvature-modulated coupling term,
$Q_0 = \alpha \kappa \rho_m \dot{\phi} \left(1 - \frac{\beta R}{6H^2} \right),$
where \( \rho_m \) is the dark matter density, \( \dot{\phi} \) is the scalar field velocity, and \( R \) is the Ricci scalar. This form of the interaction is inspired by effective field theory (EFT) considerations, where couplings between matter currents and scalar field derivatives naturally arise in the low-energy limit. The inclusion of the dimensionless curvature-modulating factor \( \beta R / 6H^2 \) introduces a geometric feedback mechanism that dynamically regulates the interaction strength depending on the cosmic epoch. At early times, when the curvature \( R \) is large, the interaction is strongly suppressed, thereby preserving standard matter-dominated evolution and allowing structure formation to proceed unimpeded. As the universe expands and \( R \) decreases, the interaction is gradually activated, enabling energy transfer from dark energy to dark matter or vice versa. This curvature-sensitive structure provides a natural and covariant mechanism to control the timing and impact of the coupling, ensuring compatibility with early-universe constraints while driving late-time acceleration without requiring fine-tuned scalar field potentials.\\

To analyze the dynamical properties of the curvature-modulated interaction model, we constructed a three-dimensional autonomous system in terms of the variables \( x, y, u \), where \( x \) and \( y \) encode the scalar field’s kinetic and potential contributions, and \( u \) denotes the logarithmic growth rate of matter perturbations. The inclusion of \( u \) as a dynamical variable is particularly significant, as it allows a unified treatment of both background expansion and structure formation within the same phase-space framework. The evolution of \( u \) directly reflects the efficiency of perturbation growth and thus provides an observable link between theoretical dynamics and large-scale structure formation. A comprehensive critical point analysis reveals a rich landscape of fixed points with distinct physical interpretations. The set includes the purely scalar-field dominated attractors \( E_{\pm} \), which feature vanishing growth and correspond to accelerated expansion with frozen perturbations; the curvature-sensitive attractors \( F_{\pm} \), which allow non-zero growth and serve as intermediates between clustering and acceleration; and the dynamically rich solutions \( J_{\pm} \) and \( I_{\pm} \), which exhibit highly nonlinear behavior due to their dependence on all three model parameters \( \alpha \), \( \beta \), and \( \lambda \).  Importantly, for both benchmark values \( \beta = 0.5 \) and \( \beta = 0.05 \), the critical points \( E_{\pm} \), \( F_{\pm} \), and \( J_{\pm} \) are found to admit stable solutions over well-defined regions in the \( \alpha\text{--}\lambda \) parameter space. These stable attractors play a pivotal role in determining the universe's asymptotic fate and structure formation history. In particular, \( F_{\pm} \) and \( J_{\pm} \) allow the system to support accelerated expansion while still accommodating non-negligible perturbation growth, making them especially relevant for reconciling cosmic acceleration with observational bounds on large-scale structure. \\

A key result of this study is the crucial role played by the curvature-modulation parameter \( \beta \) in governing the background expansion and structure growth dynamics of the interacting dark energy–dark matter model. For small values of \( \beta \), such as in benchmark (ii) with \( \alpha = 2, \lambda = 1, \beta = 0.05 \), the interaction becomes effective earlier due to weak curvature suppression, leading to enhanced energy transfer and stronger structure growth at intermediate redshifts. The phase-space evolution reveals a global late-time attractor at the critical point \( F_+ \), characterized by a non-zero perturbation growth rate \( u \neq 0 \), confirming the model’s capability to simultaneously support acceleration and structure formation. In contrast, benchmark (i) with \( \alpha = 1, \lambda = -1, \beta = 0.5 \) exhibits strong curvature screening, which delays the activation of the interaction until late times. The phase-space trajectories in this case converge to the attractor \( E_+ \), a purely scalar-field dominated solution with \( u \rightarrow 0 \), reflecting suppressed growth and consistent late-time acceleration. This dichotomy highlights the model's flexibility: while benchmark (ii) supports a dynamical regime with stronger clustering and early-time deviations from \(\Lambda\)CDM, benchmark (i) achieves a smoother transition to scalar-field dominance with suppressed early-time interaction. Importantly, in both benchmarks, the fixed points \( E_\pm \), \( F_\pm \), and \( J_\pm \) appear as viable stable attractors in the \( \alpha\text{--}\lambda \) parameter space for both values of \( \beta \), underscoring the robustness of the phase-space structure. The presence of a globally stable attractor in each scenario—\( E_+ \) for benchmark (i) and \( F_+ \) for benchmark (ii)—provides predictive control over late-time cosmic evolution and confirms the viability of curvature-modulated interactions in explaining both cosmic acceleration and structure formation within a unified dynamical framework.\\

We further analyzed the evolution of the coincidence parameter \( r_{\text{mc}} = \Omega_m / \Omega_\phi \) alongside the total effective equation of state \( \omega_{\text{tot}} \). For benchmark (i) with \( \alpha = 1, \lambda = -1, \beta = 0.5 \), the interaction remains suppressed during the matter-dominated era due to strong curvature modulation, leading to a prolonged tracking phase where \( r_{\text{mc}} \sim \mathcal{O}(1) \). This sustained coexistence of dark energy and dark matter dynamically alleviates the cosmic coincidence problem. In contrast, benchmark (ii) with \( \alpha = 2, \lambda = 1, \beta = 0.05 \), characterized by weaker curvature suppression and stronger coupling, activates the interaction earlier. This results in a sharper drop in \( r_{\text{mc}} \), signaling a rapid transition to scalar-field domination. The color gradient in the plot encodes the evolution of \( \omega_{\text{tot}} \), indicating that both models achieve late-time acceleration while exhibiting qualitatively different intermediate histories.  We have also examined the effective total equation of state $\omega_{\rm tot}$ for both benchmark cases in fig.\ref{fig:5} and fig.\ref{fig:6}. In neither case does $\omega_{\rm tot}$ cross the phantom divider ($\omega_{\rm tot} = -1$). Instead, the models approach stable late-time acceleration with $\omega_{\rm tot}$ asymptoting to values in the non-phantom regime for the redshift range $0 < z < 2.3$. Thus, our curvature-modulated interaction provides viable late-time acceleration without requiring phantom dynamics.\\

 In the perturbative sector, we analysed the evolution of four key growth observables—the growth rate of matter perturbations $u(N)$, the matter density contrast $\delta_m(z)$, the growth index $\gamma(z)$, and the combined diagnostic $u\sigma_8(z)$—across two benchmark scenarios. Benchmark (i), with $(\alpha = 1, \lambda = -1, \beta = 0.5)$, represents moderate coupling with stronger curvature modulation, while benchmark (ii), with $(\alpha = 2, \lambda = 1, \beta = 0.05)$, corresponds to stronger coupling and weaker curvature suppression. Including both the previously compiled redshift-space distortion (RSD) measurements and the recent DESI DR1 RSD data points, we find that benchmark (ii) shows markedly enhanced growth at earlier times. This appears as a sharp peak in the growth rate $u(N)$ near $N\sim1$, indicating a period of efficient structure formation and a correspondingly larger $\delta_m(z)$ over the full redshift range. The growth index $\gamma(z)$ for this case falls from $\sim0.57$ to below $0.4$ by $z\sim2$, signalling a faster transition from growth to acceleration, and the predicted $u\sigma_8(z)$ tracks the observed rise and subsequent decline around $z\sim0.8$, remaining fully consistent with both the older RSD compilation and the DESI DR1 constraints. By contrast, benchmark (i) produces a flatter growth profile, with a suppressed $u(N)$ during the matter-dominated era and a delayed interaction onset, keeping $\delta_m(z)$ lower and $\gamma(z)$ closer to its $\Lambda$CDM value of $\gamma\simeq0.55$. Although benchmark (i) yields a less pronounced peak in $u\sigma_8(z)$, it also stays within the combined RSD and DESI DR1 observational envelope. Together, these results underscore how the inclusion of both legacy RSD and DESI DR1 RSD data sharpens the distinction between curvature-modulated interacting scenarios and $\Lambda$CDM: benchmark (ii) signals enhanced early-time clustering and efficient growth, whereas benchmark (i) represents a delayed, conservative growth history compatible with late-time acceleration. In this way, the combined observables $\gamma(z)$ and $u\sigma_8(z)$ emerge as powerful diagnostics for testing curvature-modulated interactions with forthcoming survey data.\\

 To evaluate the compatibility of our interacting dark energy–dark matter framework with observational data, we studied the evolution of background observables such as the Hubble parameter $H(z)$ and the distance modulus $\mu(z)$, together with cosmographic quantities like the deceleration parameter $q(z)$ and the jerk parameter $j(z)$. Both benchmark scenarios—(i) $(\alpha = 1, \lambda = -1, \beta = 0.5)$ and (ii) $(\alpha = 2, \lambda = 1, \beta = 0.05)$—closely follow the $\Lambda$CDM background evolution and remain within current observational bounds. In particular, the $H(z)$ curves for both benchmarks were compared with OHD measurements and DESI + QSO + Ly$\alpha$ forest data, showing nearly indistinguishable behaviour from $\Lambda$CDM at low to intermediate redshifts, with only mild deviations emerging at higher redshifts ($z \gtrsim 1.5$), especially for benchmark~(ii) due to its earlier activation of the interaction. Likewise, the distance modulus $\mu(z)$ was tested against Pantheon+ supernova data combined with DESI + QSO + Ly$\alpha$ forest data, confirming that both models deliver excellent fits throughout the redshift range $0 < z < 2.3$, with benchmark~(ii) slightly outperforming benchmark~(i) at the high-$z$ end thanks to its sharper transition to scalar field domination. Beyond background dynamics, we examined the cosmographic evolution encoded in $q(z)$ and $j(z)$. Both deceleration and jerk parameter plots reveal a smooth transition from deceleration ($q > 0$) to acceleration ($q < 0$), consistent with observational expectations. However, benchmark~(ii), due to its stronger coupling and weaker curvature suppression, transitions to acceleration slightly earlier than benchmark~(i), demonstrating enhanced responsiveness to background evolution. The right panel shows the jerk parameter $j(z)$, which is identically unity in $\Lambda$CDM. Benchmark~(i) exhibits only modest deviation from $j=1$, whereas benchmark~(ii) displays a pronounced peak in $j(z)$ at intermediate redshifts, signalling non-trivial dynamics from the interaction term. These deviations reflect the underlying scalar field kinetics and the time-dependent activation of the coupling, governed by the curvature-modulated exponential suppression factor. The more substantial variation in $j(z)$ for benchmark~(ii) underscores the potential of this cosmographic observable as a sensitive probe of new physics in future high-precision surveys. Altogether, these results affirm that the interacting model remains observationally consistent at the background level, while the cosmographic parameters—particularly $j(z)$—encode unique imprints of the underlying curvature-modulated interaction. This dual agreement with current OHD, DESI + QSO + Ly$\alpha$ forest, and Pantheon+ datasets, together with the potential for detectability of $j(z)$, makes both benchmarks compelling alternatives to $\Lambda$CDM and motivates future investigations with forthcoming surveys.\\

 This work offers a significant advancement in modeling interacting dark energy through a curvature-modulated framework, where the interaction strength is dynamically controlled by the Ricci scalar via the parameter $\beta$. The exponential factor $e^{-\gamma R}$, governed by $\beta$, acts as a natural screening mechanism: it suppresses the interaction at high curvature, preserving early-universe physics and structure formation, while dynamically activating at late times to drive accelerated expansion and alleviate the coincidence problem. Derived from an interaction Lagrangian, the model ensures covariant conservation, embeds naturally in an EFT-inspired framework, and enables a unified, self-consistent treatment of both background and perturbation dynamics—distinguishing it from many purely phenomenological $Q$ forms. The ability to regulate cosmic evolution by tuning the parameters \( \alpha \) and \( \lambda \) further enhances the model’s flexibility, allowing smooth transitions between early-time matter domination and late-time scalar-field domination. The presence of stable critical points across wide regions of the parameter space underscores the robustness and dynamical richness of the model. From a phenomenological standpoint, the model successfully reproduces key cosmological features: alleviation of the coincidence problem, agreement with observational growth data, and mild deviations in cosmographic parameters like \( j(z) \), which can serve as potential observational signatures. These features make the framework a strong alternative to \( \Lambda \)CDM and highly relevant for current and future precision cosmology. Looking ahead, this model sets the stage for a number of future explorations. These include full MCMC-based parameter constraints, detailed comparison with weak lensing and redshift-space distortion data, and investigations into non-linear structure formation. Its ability to dynamically couple dark sectors with geometric modulation opens new avenues for understanding cosmic acceleration in a theoretically consistent and observationally testable way, positioning it as a compelling candidate for next-generation cosmological probes such as \textit{Euclid}, \textit{DESI}, and \textit{LSST}.

 \vspace{1cm}
\paragraph*{Acknowledgement}
This work is supported in part by the National Key Research and Development Program of China under Grant No. 2020YFC2201504. \\
 
\paragraph*{Data Availability Statement} All original datasets used in this work are publicly accessible and have been appropriately cited in the bibliography.

\end{document}